# Few-layer Tellurium: one-dimensional-like layered elementary semiconductor with striking physical properties

Jingsi Qiao[1,2§], Yuhao Pan[1§], Feng Yang[1§], Cong Wang[1], Yang Chai[2] and Wei Ji[1, *]

[1]*Beijing Key Laboratory of Optoelectronic Functional Materials & Micro-Nano Devices, Department of Physics, Renmin University of China, Beijing 100872, China*

[2]*Department of Applied Physics, The Hong Kong Polytechnic University, Hong Kong, China*

**Abstract**

Few-layer Tellurium, an elementary semiconductor, succeeds most of striking physical properties that black phosphorus (BP) offers and could be feasibly synthesized by simple solution-based methods. It is comprised of non-covalently bound parallel Te chains, among which covalent-like feature appears. This feature is, we believe, another demonstration of the previously found covalent-like quasi-bonding (CLQB) where wavefunction hybridization does occur. The strength of this inter-chain CLQB is comparable with that of intra-chain covalent bonding, leading to closed stability of several Te allotropes. It also introduces a tunable bandgap varying from nearly direct 0.31 eV (bulk) to indirect 1.17 eV (2L) and four (two) complex, highly anisotropic and layer-dependent hole (electron) pockets in the first Brillouin zone. It also exhibits an extraordinarily high hole mobility (~$10^5$ cm$^2$/Vs) and strong optical absorption along the non-covalently bound direction, nearly isotropic and layer-dependent optical properties, large ideal strength over 20%, better environmental stability than BP and unusual crossover of force constants for interlayer shear and breathing modes. All these results manifest that the few-layer Te is an extraordinary-high-mobility, high optical absorption, intrinsic-anisotropy, low-cost-fabrication, tunable bandgap, better environmental stability and nearly direct bandgap semiconductor. This "one-dimension-like" few-layer Te, together with other geometrically similar layered materials, may promote the emergence of a new family of layered materials.

---

\* wji@ruc.edu.cn
§ contributed equally to this work

# 1. Introduction

Two-dimensional (2D) layers have been demonstrated a marvelous category of materials functionalized as electronic and optoelectronic devices, e.g. transistors [1-4], photo detectors [5, 6], light absorbers [4, 7], memories [8], switcher [9]. Few-layer black phosphorus (FLBP), isolated from its bulk from in 2014, was evidenced an extremely promising candidate for further electronics [3, 4, 6, 10]. It offers moderate electronic and direct optical bandgaps [4], high mobility [3, 4, 10], optical linear dichroism [4] and among the other striking properties [11, 12], which bridges the gap between graphene and transition metal dichalcogenides (TMDs). Two remanding issues of FLBP lie in the effective protection and high efficient preparation of FLBP. The protection was recently realized by capping with addition 2D layers [13], metal-oxides or chemically attached radicals or molecules [14]. The fabrication of BP is, however, in huge challenge to the community that the growth of bulk BP needs high-pressure and the bottom-up synthesis of FLBP has yet to be achieved, in contrast to the situation of graphene and TMDs.

In light of this, it calls for a new material that is much easier to synthesize, with better environmental stability and succeed, at least most of, those striking properties of FLBP, namely moderate bandgap, high electronic carrier mobility and strong optical absorption, anisotropy and ideally direct bandgap. Few-layer Tellurium is a candidate of such material. The solution-based synthesis method of it, by mixing two or more chemicals in a flask [15, 16], extremely lowers the cost and simplifies the procedure for preparation. In addition, it is a layered material comprised of quasi-one-dimensional chains, which introduces intrinsically geometric anisotropy in the layer plane since the atoms are bonded by covalent and non-covalent bonds in the two directions, respectively. Although the bulk, nanoplate and nanowire forms of it were demonstrated of interest in topological materials [17], thermoelectric [18-21], optoelectronic [22] and piezoelectric [23, 24] applications, it is expected that emerging striking properties in its layered (2D) form are yet to be explored.

Here, we focus on the theoretical prediction of properties of few-layer α-Te (FL-α-Te) as it is always the most stable phase if the thickness beyond 1L. The FL-α-Te, similar to BP, has a tunable moderate bandgap varying from 1.17 eV (2L) to 0.31 eV (bulk) with inclusion of spin-orbit coupling (SOC). It is interesting that the maps of the valence band (VB) and conduction band (CB) surfaces in the *k*-space substantially develop from 2L to bulk, leading to complex and anisotropic electron and hole pockets. Both the tunable bandgap and the evolutionary VB and CB surfaces are stemmed from the formed covalent-like features among the inter-chain Te atoms, which is, we believe, another demonstration of the covalent-like quasi-bonding (CLQB) found in BP [4, 11], $PtS_2$ [25, 26] and $PtSe_2$ [27]. It is exceptional that FL-α-Te offers an extraordinarily high hole-mobility, ~$10^5$ cm$^2$/Vs, and strong optical-absorption in the visible (up to 9% per layer) and infrared light regions in the CLQB direction. The CLQB inter-chain coupling also yields nearly isotropic light absorption from a highly anisotropic geometry, anomalous interlayer vibrational properties and comparably high force constants among the all three directions. It is rather striking that the CLQB direction, a non-covalently bound direction, offers a much higher carrier mobility and a reasonably larger force constant than the covalently connected direction. FL-α-Te is highly stretchable that the ideal strength of it is up to 18%–26% while its bulk form shows 26%–38%. Beyond those ranges, the Te chain does not breakdown but new phases emerge. A rather high oxidization resistance of FL-α-Te was found, suggesting its good environmental stability. All these results manifest FL-α-Te is an extraordinary-high-mobility, high optical absorbance, intrinsic anisotropy, low-cost fabrication, tunable bandgap, nearly direct bandgap and "1D-like" layered semiconductor, which may bring emerging properties and opportunities for the research of layer materials.

## 2. Methods

## 2.1. DFT calculation

Density functional theory calculations were performed using the generalized gradient approximation for the exchange-correlation potential, the projector augmented wave method [28, 29] and a plane-wave basis set as implemented in the Vienna *ab-initio* simulation package (VASP) [30] and Quantum Espresso (QE) [31]. Density functional perturbation theory was employed to calculate phonon-related properties, including Raman intensity (QE), activity (QE) and shifts (VASP), vibrational frequencies at the Gamma point (VASP) and other vibration related properties (VASP). The kinetic energy cut-off for the plane-wave basis set was set to 700 eV for geometric and vibrational properties and 300 eV for electronic structure calculations. A $k$-mesh of 1×15×11 was adopted to sample the first Brillouin zone of the conventional unit cell of few-layer Te in all calculations. The mesh density of $k$ points was kept fixed when calculating the properties for bulk Te. In geometric optimization and vibrational property calculations, van der Waals interactions were considered at the vdW-DF [32, 33] level with the optB88 exchange functional (optB88-vdW) [34-36], which was proved to be accurate in describing the structural properties of layered materials [4, 37, 38]. The theoretical lattice constants of few-layer and bulk α-Te were double checked with the optB86b-vdW [34, 35], optPBE-vdW [34-36], SCAN [39, 40] and SCAN-rVV10 functionals [39, 40], as available in Table S1 (online). Although SCAN offers the best lattice constants, but the HSE06 bandgap calculated based on the SCAN structure is roughly 0.2 eV too large while the value based on the optB88-vdW structure perfectly fits the experimental value. Given the second best functional for predicting lattice constants, we adopted the optB88-vdW functional for our calculations of structure-related properties in this work. The shapeand volume of each supercell were fully optimized and all atoms in the supercell were allowed to relax until the residual force per atom was less than $1\times10^{-4}$ eV·Å$^{-1}$. Electronic bandstructures were calculated using the optB88-vdW functional and the hybrid functional (HSE06) [41, 42] with and without spin-orbit coupling (SOC). Surface maps of valence and conduction bands shown in Section 3.2 were calculated using the optB88-vdW functional.

## 2.2. Carrier mobility prediction

In 2D Te, the carrier mobility is estimated by [43]

$$\mu_{\text{film}} = \frac{\pi e \hbar^4 C_{\text{film}}}{\sqrt{2}(k_B T)^{3/2}(m^*)^{5/2}(E_1^i)^2} \cdot F. \quad (1)$$

Here, $F$ is a crossover function that bridges 2D and 3D, where

$$F \equiv \frac{\sum_n \{\frac{\sqrt{\pi}}{2}[1 - erf(\Omega(n))] + \Omega(n)e^{-\Omega^2(n)}\}}{\sum_n [1 + \Omega^2(n)]e^{-\Omega^2(n)}}$$

and $\Omega(n) \equiv \frac{n\pi\hbar}{\sqrt{2m^* W_{\text{eff}}^2 k_B T}}$.

$m^*$ is the effective mass along the transport direction. $W_{\text{eff}}$ is the effective thickness of the film, which is analyzed by

$$\frac{1}{W_{\text{eff}}} = \int_{+\infty}^{-\infty} P_i(x) P_f(x) dx = \sum_n \frac{\rho_i^n(x)}{\sum_n \rho_i^n(x) \Delta x} \frac{\rho_f^n(x)}{\sum_n \rho_f^n(x) \Delta x} \Delta x = \frac{1}{N^2 \Delta x} \sum_n \rho_i^n(x) \rho_f^n(x).$$

Therein, $P(x)$ is the electron probability density along the $x$ direction. We divided the space along the $x$ direction into $n$ parts by $\Delta x$. Variable $\rho^n(x)$ is the number of electrons from $(n-1)\Delta x$ to $n\Delta x$ along the $x$ direction, which could be calculated by solving the DFT equations. Here, $N$ is the total number of valence electrons in the film, $i$ and $f$ represent equilibrium and deformed films, respectively. Term $E_1$ represents the deformation potential constant of the valence band maximum (VBM) for hole or conduction band minimum (CBM) for electron along the transport direction ($y$ or $z$), defined as $E_1^i = \Delta V_i / (\Delta l / l_0)$. Here $\Delta V_i$ is the energy change of the $i^{\text{th}}$ band under uniaxial compression and dilatation (by a step 0.25 % or 0.5%), $l_0$ is the corresponding lattice constant along the transport direction and $\Delta l$ is the deformation of lattice constant. The estimation of error for $E_1$ is summarized in the Supplementary data (online). $C_{\text{film}}$ is the elastic modulus of the longitudinal strain in the propagation direction, which is derived by $(E-E_0)/V_0 = C(\Delta l/l_0)^2/2$, where $E$ is the total energy and $V_0$ is the effective lattice volume at the equilibrium for 2D systems, $V_0 = S_0 \cdot W_{\text{eff}}$. The temperature for all mobility calculations is 300 K. The carrier mobility is also evaluated by the well-known 2D formula [4, 37]

$$\mu_{2D} = \frac{e\hbar^3 C_{2D}}{k_B T m_e^* m_d (E_1^i)^2}. \qquad (2)$$

The details for comparison of these two formulas are shown in Table S2 (online).

### 2.3. Optical absorption spectra calculation

Absorption spectra were calculated from the dielectric function using the expression for few-layer Te, $A(\omega) = \alpha(\omega)*\Delta x$, where $\alpha(\omega) = \frac{\omega \varepsilon_2}{cn}$ is the absorption coefficient, $n = \sqrt{\frac{\sqrt{\varepsilon_1^2 + \varepsilon_2^2} + \varepsilon_1}{2}}$ is the index of refraction, $\varepsilon_1$ and $\varepsilon_2$ are the real and imaginary parts of the dielectric function [44], $\omega$ is the light frequency, $c$ is the speed of light *in vacuo* and $\Delta x$ represents the unit-cell size in the $x$ direction. The electronic structures were obtained from the results unveiled using the optB88-vdW functional with SOC and the $k$-mesh was increased to 1×29×25 in calculating dielectric functions. Enough conduction bands were considered and exciton effects were not considered in the optical properties calculations. Because the dielectric function is a tensor, the absorption spectra along the $x$, $y$ and $z$ directions were obtained separately. The energies of incident light of the horizontal axis in absorption spectra (Section 3.5) were shifted by the differences of bandgaps between the optB88-vdW and HSE06 results with the inclusion of SOC.

### 2.4. Calculation of force constant

In a rigid layer vibrational mode, the whole layer can be treated as one rigid body [25, 45]. The projected interlayer force constant (p-ILFC) $k^i$ ($i$ stands for the projected direction, e.g. $x$, $y$ or $z$) was constructed by summing inter-atomic force constants over all atoms from each of the two adjacent layers, as $k_{AB}^i = \sum_{a,b} D_{a,b}^i$ ($a \in$ [atoms in layer $A$], $b \in$ [atoms in layer $B$]). The matrix of inter-atomic force constants, essentially the same as the Hessian matrix of Born-Oppenheimer energy surface, is defined as the energetic response to a distortion of atomic geometry in DFPT [46]. It reads, $D_{a,b}^i = \frac{\partial^2 E(\mathbf{R})}{\partial R_a^i \partial R_b^i}$, where $\mathbf{R}$ is the coordinate of each ion, $E(\mathbf{R})$ is the groundstate energy. The force on an individual ion can thus be expressed to $F_i = -\frac{\partial E(\mathbf{R})}{\partial \mathbf{R}_i}$.

Vibrational frequency $\omega$ are related to the eigenvalues of the Hessian matrix and atomic masses. It reads $\det \left| \frac{1}{4\pi^2 c^2} \frac{D_{a,b}}{m_{a,b}} - \omega^2 \right| = 0$, where $m_{ab}$ is the effective mass, defined as $\sqrt{m_a m_b}$.

**2.5. Evaluation of environmental stability**

The adsorption energies of $O_2$ and $H_2O$ on 50 and 52 initial sites of a 4 ($y$) ×3 ($z$) Te bilayer supercell were considered. All calculations, including the adsorption energy and oxidization barrier estimations, were performed using the optB88-vdW functional. The adsorption energies of the physi- and chemi-sorbed $O_2$ configurations were double checked using the SCAN-rVV10 functional, while the bonding energy of $O_2$ were compared using the optB88-vdW, SCAN [39] and SCAN-rVV10 functionals as shown in Table S3 (online). The kinetic energy cutoff was set to 400 eV and a $k$-mesh of 1×3×3 was used to sample the surface BZ. The shape ($y$-$z$ plane lattice parameters) of each supercell was fixed and all atoms in the supercell were allowed to relax until the residual force per atom was less than 0.02 eV·Å$^{-1}$. The potential-energy profiles were calculated using the climbing-image nudged elastic band (CI-NEB) technique [47] for the reaction pathways from physisorbed $O_2$ to dissociative-chemisorbed $O_2$ on a pristine Te bilayer, which locates the exact saddle points of the pathways. This method was proofed an effective way of estimating oxidation barriers in few-layer BP [48, 49].

# 3. Results and discussion

**3.1. Geometry and stability of Tellurium allotropes**

The α-phase is the only normal-pressure phase for bulk Tellurium (Fig. 1a, space group $P3_121$ (No. 152)) [50-52]. It is comprised of parallel-aligned Te triangle chains, in which each repeating unit contains three Te atoms. If the thickness of the α-phase few layers (Fig. 1b–1d, space groups $P2$ (No.3) for odd and $P2_1$ (No.4) for even numbers of layers) reduces to the monolayer limit, the β-phase (Fig. 1e and 1f, space groups $P2/m$ (No. 10) for odd and $P2_1/m$ (No. 11) for even numbers of layers) emerges (see Fig. S1 online). The both phases share most of features in the few-layer form, but a Te atom of a triangle chain moves towards the adjacent chain in the β-phase, resulting in an additional mirror symmetry (from Fig. 1c ($P2$) to 1f ($P2/m$)). The β-phase is more stable for the monolayer where the α-phase cannot sustain and transfers to the β-phase without a barrier. However, this transformation does not occur for the Selenium case where the α-phase is still the most stable monolayer phase. Another way to reduce the dimension of bulk α-Te lies in cutting the Te-Te covalent bonds along the *x-y* plane, which introduces dangling bonds to the cut few-layers and is thus more difficult to realize. However, if it could be achieved, a phase transition to the γ-phase (Fig. 1g and 1h, space group $P\bar{3}m1$ (No. 164)) occurs for lowering the energy of dangling bonds when the number of Te atoms in each chain is less than nine atoms and could be divided by three. The γ-phase is a variation to the α-phase in the *x-z* plane, in analogy to the 1T phase of TMDs, see Fig. S1 (online). This transition is a result of the subtle balance of the total energies between the two phases. The critical number of atoms for the transition may increase if stronger surface-layer attraction is introduced to the dominant phase after the transition.

If the layer thickness is larger than 1L, the *α*-phase is, however, always more stable than other phases at equilibrium, as shown in Fig. 1i. The γ-phase completely transforms into the *α*'-phase at 5L, which stands for the 90°-rotated *α*-phase (Fig. S1 online). However, this phase is less stable than either the *α*- or β-phase. The energy difference between the *α*- and the second-lowest-energy (γ for 2L and *β* for thicker layers) phases is only 1 meV per Te for 2L and increases to 8 meV for 3L, which are comparable with the difference of 2 meV between black and blue phosphorus [53]. The phonon dispersion spectra of 2L β-Te does contain a significant imaginary frequency (not shown here) representing the motion of transferring from the β to α phase. The energy difference enlarges to 11 meV for 4L and is prone to converge to 18 meV for bulk. These results thus compellingly indicate the stability of the α phase in 2L and thicker layers. A reversible *α-β* or *α-γ* phase transition may occur with in-plane strain or charge doping applied due to the small energy difference between them and their geometrical similarity. In particular, the Te few-layer favors the α-phase under biaxial expansion to the lattice but prefers the β-phase for compression (Fig. S2a online). In terms of uniaxial strain, either compression along both the in-plane directions or expansion along the *z* direction could lead to an α to β phase transition (Fig. S2b, S2c online). Our primary results also suggest charge doping induced phase transitions occurred between the α and β phases. It also suggests that additional substrate-layers may maintain the meta-stable (β- or γ-) phase in 2L or thicker layers owing to the substrate induced in-plane strain or charge transfer [54]. Given the better stability of it, here, we solely focus on the α-phase.

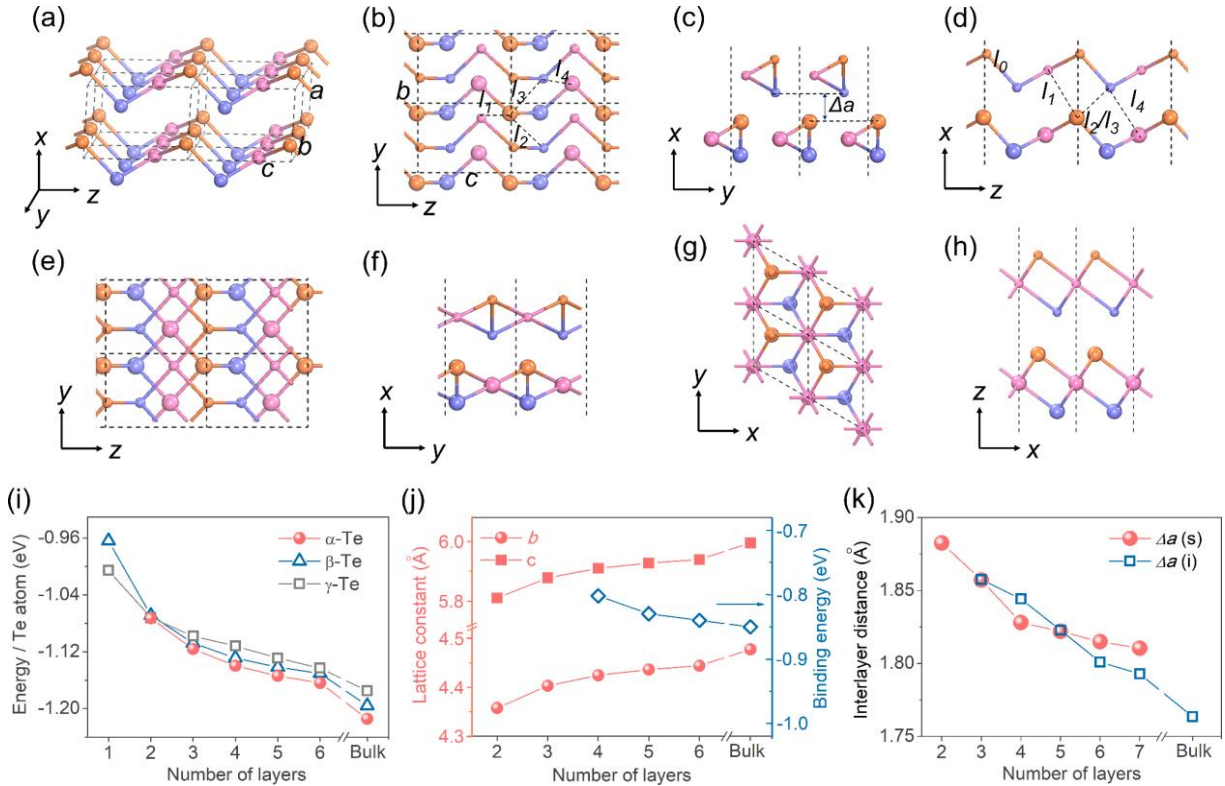

Fig. 1. (Color online) Geometry and stability of Te allotropes. (a)–(d) Crystal structures of bulk (a) and bilayer α-Te in the top- (b) and side-views ((c) and (d)). (e)–(f) Crystal structures of bilayer β-Te in the top- (e) and side-views (f). (g)–(h) Crystal structures of bilayer γ-Te in the top- (g) and side-views (h). (i) Layer-dependent energy stability of α-, β- and γ-Te. (j) Lattice constants (*b* and *c*) and binding energy of few-layer and bulk α-Te. (k) Surface (s) and inner (i) interlayer distances, $\mathit{\Delta}a$, as marked in (c), of few-layer and bulk α-Te.

Fig. 1b–1d shows the top and side views of FL-α-Te. The surface of the layer is in the *y-z* plane and the lattice constants are denoted *b* and *c*, respectively. Explicitly layer-dependence for both *b* and *c* are shown in Fig. 1j that *b* increases from 4.36 Å of 2L to 4.44 Å of 6L and 4.48 Å of bulk. Constant *c* shares the same tendency with *b* varying from 5.81 to 5.99 Å. This variation implies potentially strong interlayer interaction in FL-α-Te, which is partially confirmed by the exceptionally large interlayer binding energy of –0.80 eV/unit-cell for 4L (–0.85 eV/unit-cell for bulk). This energy is nearly twice to that of BP which was proved to have strong interlayer coupling [4]. Each Te chain in a layer has two nearest neighboring chains in the adjacent layer. The intra-layer bond length $l_0$ is in a range from 2.83 to 2.92 Å. Each Te atom in a chain has maximum three nearest neighboring Te atoms in the two neighboring chains with inter-atom distances $l_1$ to $l_4$ varying from 3.47 to 3.56 Å (Fig. 1b and 1d). The neighboring number reduces to two, with the distances of 3.41 Å, in Te double-chains as shown in Fig. S3b (online). The 0.8 eV binding energy is thus straightforward if given the inter-chain binding energy of 0.41 eV for a double chain (Fig. S3 online). The slightly increased interlayer binding energy (Fig. 1j) is accompanied by the shortened interlayer distance (Fig. 1k) with respect to layer thickness.

### 3.2. Electronic bandstructures of few-layer Te

Our calculated bandstructures show bulk Te is a nearly direct-bandgap semiconductor with the bandgap of 0.31 eV near the H point (0.33 0.33 0.50) in the irreducible Brillouin zone (iBZ) (Fig. 2a and 2c), highly consistent with the experimental value of 0.33–0.34 eV [55, 56]. The VBM and CBM locate at (0.33 0.33 0.485) and (0.33 0.33 0.495) in the iBZ, respectively. Spin-orbit coupling (SOC) plays an essential role in predicting the bandgap that the inclusion of it reduces the bandgap by 0.20 eV. The thinnest few-layer Te in the α-phase is 2L. Fig. 2d shows its bandstructure, indicating an indirect bandgap of 1.17 eV. The bandstructures of 3L to 6L are available in Fig. S4 (online) and the coordinates of VBM and CBM of 2L to bulk are listed in Table S4 (online). The valence band edge of 2L-α-Te appears to reside at a point between S and G (point H', the projection of H in the 2D iBZ, see Fig. 2b) and CBM at Z (Fig. 2d). Given mappings of the conduction and valence bands in the 2D BZ (Fig. 2e and f), the position of VBM for 2L was found not locating at the H' (0.33, 0.50) point, but a point near it and off the S-G line (±0.27, ±0.34) (Fig. 2f). There are thus four valleys in the first BZ for VB, which are, most likely, (pseudo) spin degenerated due to strong SOC in Te. Here, we do not plot the mappings in the first BZ along $k_z$, but from 0.0 to 1.0 (2π/c) of $k_z$ for better showing the evolution of VBM and CBM from 2L to bulk. Mappings plotted in the first BZ are available in Fig. S5 (online). These results suggest a "M-shaped" VB for 2L-α-Te (noted with the white dashed line), which were usually found in topological insulators with VB-CB band inversion occurred and play a key role for high-performance thermoelectrics. Indeed, both positions of VBM and CBM do move in from 2L to 4L (Fig. 2g and 2h) and 6L (Fig. 2i and 2j). Those of 3L and 5L are available in Fig. S6 (online). The "M-shaped" VB still holds for 6L although the complicity, size and anisotropy reduces (white lines noted in Fig 2f, 2h and 2j). There are two valleys for CBM in the 2D iBZ, one at (0.0, 0.27) and the other very close to H' at (0.27, 0.50) (Fig. 2e) in 2L. The former is deeper in 2L and both of them are nearly degenerated for 3L, but the latter becomes the real CBM for thicker layers (Fig. S6 (online), 2g, 2i and 2k), e.g. 6L (Fig. 2i). However, even for 6L, the position of VBM is still appreciably off the H' point and it eventually approaches H in bulk Te (Fig. 2l). The even stronger anisotropy and larger complicity of Te few-layers for the electron and hole pockets imply potentially even better performance of FL-α-Te than that of Te nanowire in thermoelectrics. The 8 meV energetic difference between the direct and indirect bandgaps of the bulk form is rather small compared with the thermal excitation energy at room temperature, although it enlarges to tens of meV for few-layers. It thus suggests that thermal-activated optical transitions between the CB and VB may occur for thicker layers, which likely leads to a bright-to-dark transition of optical signals at a certain temperature.

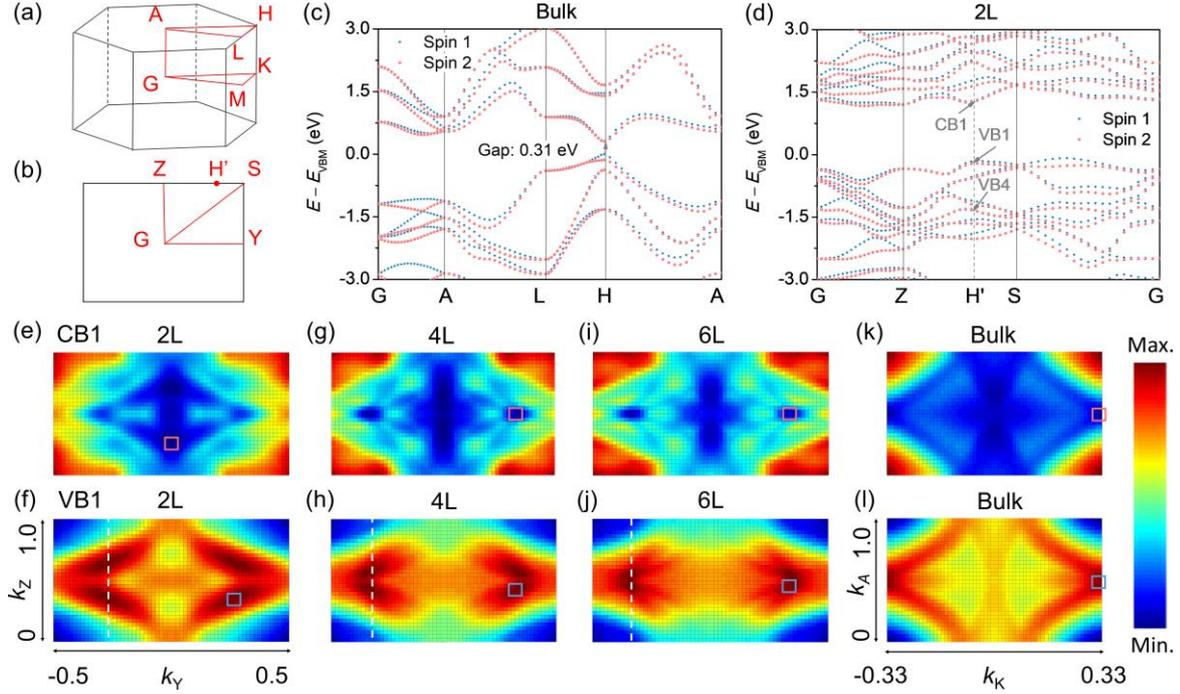

Fig. 2. (Color online) Electronic bandstructures of few-layer Te. (a), (b) Brillouin zones of bulk and few-layer α-Te. (c)–(d) Bandstructures of bulk (c) and bilayer (d) α-Te. (e)–(l) Schematic drawings of energy surface for the lowest conduction and highest valence bands of 2L ((e), (f)), 4L ((g), (h)) and 6L ((i), (j)) α-Te in the 2D Brillouin Zone (b), and those of bulk α-Te ((k), (l)) in the GKHA plane of the 3D Brillouin Zone (a). Colors "red" and "blue" represent the maximum and minimum energy values, respectively.

### 3.3. Interlayer couplings in FL-α-Te

Given the surfaces of CB and VB explored, we plotted the bandgaps as a function of layer thickness in Fig. 3a. Both the vdW-DF and hybrid (HSE06) functionals were considered with and without SOC. The layer-dependent bandgap evolutions predicted using these functionals/methods essentially share the same tendency. The bandgap of 2L was predicted 1.17 eV using HSE06+SOC. It reduces to 0.95, 0.83, 0.72 and 0.66 eV for 3L to 6L, showing a slightly perturbed inverse decay as a function of the number of layers ($n_L$), which was deuced using an one-dimensional quantum well model [57] (Supplementary Note 1 and Fig. S7 online). Fig. 3b shows the evolution of CBM and VBM as a function of the number of layers. The VBM significantly changes from –4.98 to –4.35 eV, roughly three times to that of the CBM. The variation range of VBM suggests p-type contact of FL-α-Te to most of metal electrodes.

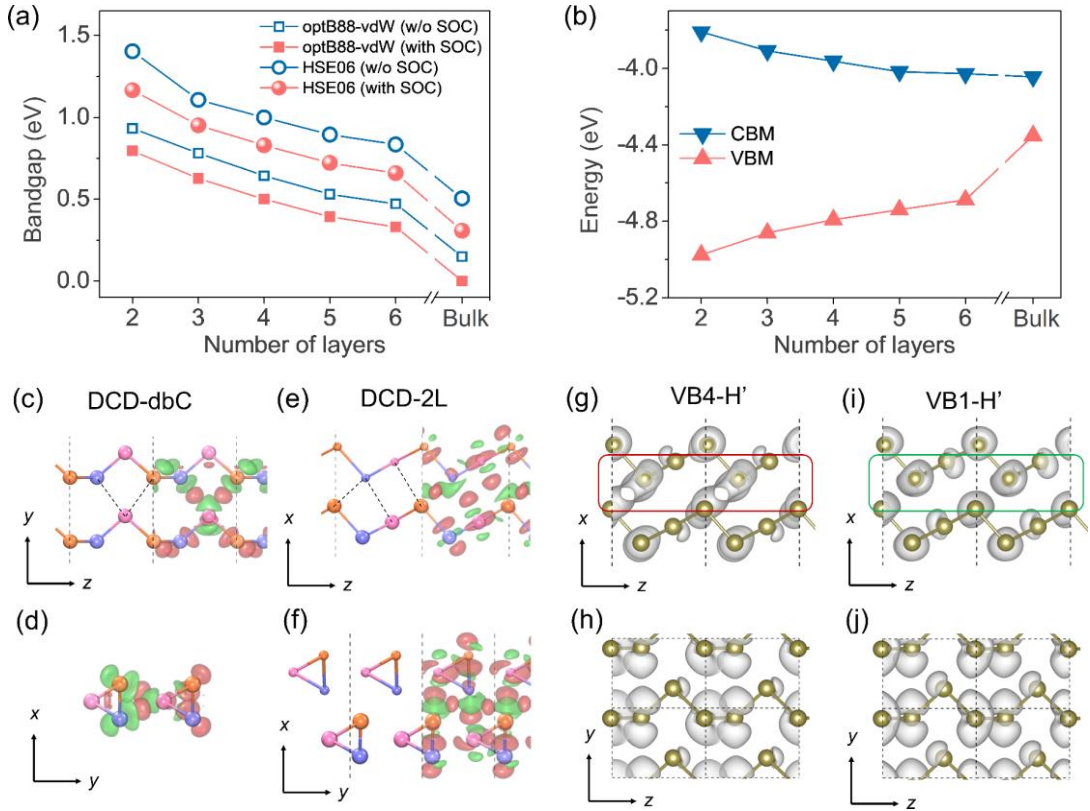

Fig. 3 (Color online) Interlayer couplings of FL-α-Te. (a), (b) Evolution of the indirect bandgaps and the positions of VBM and CBM as a function of sample thickness. (c), (d) Differential charge density of double Te atomic chains (DCD-dbC, (c), (d)) and bilayer α-Te (DCD-2L, (e), (f)). Pristine atomic structures are illustrated in left side for comparison. (g)–(i) Visualized wavefunctions for the labeled states in Fig. 2d, VB4 ((g), (h)) and VB1 ((i),(j)), of bilayer α-Te along the *x-z* and *x-y* planes using an isosurface of 0.0025 $e$ Bhor$^{-3}$.

Inter- and intra-layer couplings between Te chains share the same features since FL-α-Te is comprised of helical chains. Fig. 3c and 3d clearly shows the isosurface of differential charge density (DCD) between two Te chains in an α-Te double-chain (dbC). Along lines $l_1$ and $l_2/l_3$ marked in Fig. 1b and 1d, it explicitly shows charge reduction near the Te atoms and charge accumulation between each two Te atoms, indicting covalent-like feature of these interactions. In 2L, it increases to three that the number of these covalent-like features initialized from the blue-colored (top layer) or orange-colored (bottom layer) Te atoms, as shown in Fig. 3e and 3f. Partial charge densities, reflecting spatial distribution of wavefunctions, of 2L-α-Te are shown in Fig. 3g–3j and S6e (online). States VB4 and VB1 at the H' point were found interlayer bonding (rectangle in Fig. 3g) and anti-bonding (rectangle in Fig. 3h) states, respectively. This type of bonding is, again, covalent-like quasi-bonding since both bonding and anti-bonding states are fully occupied, in analogy to those revealed in BP [4], $PtS_2$ [25, 26] and $PtSe_2$ [27]. The decreased number of CLQB from 2L (4/3Te) to 1L (2/3Te) explains the reason why the α phase tends to transform into the β phase forming additional 2/3 covalent bonds per Te in 1L.

### 3.4. Carrier mobility of FL-α-Te

Few-layer α-Te is partially of 1D nature and these chains are bound through our revealed CLQB, but generally believed vdW, interactions, since covalent feature does emerge. In light of this, its carrier mobility might be highly anisotropic and of interest to explore. Table 1 summarizes the results of predicted phonon-limited carrier mobilities for 2L to 6L α-Te layers and compares them with previously predicted values for other high mobility 2D materials, namely BP [4] and $PtSe_2$ [27]. As shown early, 4L appears a critical thickness for electronic structures since the VBM and CBM are nearly located at the positions of bulk in 4L and thicker layers. We therefore focus on the results from 4L to 6L.

Although the Te chains appear relatively weakly bonded through CLQB along the *y* direction, the group velocities (effective masses) of both electron and hole of this direction are exceptionally comparable with those via real covalent bonds (along the *z* direction). Table 1 shows that the electron effective mass along the *y* direction varies from 0.23 to 0.28 $m_0$ while 0.30 to 0.40 $m_0$ for that of hole. Given the weaker CLQB interaction compared with covalent bond, the in-plane deformation potential along the $y$ direction ($E_{1y}$) is thus expected small, which was confirmed by our calculations. The deformation potentials were found in a range of –0.12 to 0.20 eV for hole in 5L and 6L, being at least an order of magnitude smaller to the corresponding values along the $z$ direction. Both of these striking properties of FL-α-Te lead to extraordinarily large hole mobilities along the $y$ direction, i.e. $10^4$–$10^6$ cm$^2$/Vs, in 5L and 6L, which are 1–2 orders of magnitude larger than the largest predicted hole mobility of few-layer BP (~$10^3$ cm$^2$/Vs) [4]. The predicted electron mobilities ($10^3$–$10^4$ cm$^2$/Vs) are generally 1–2 orders of magnitude smaller than the hole mobilities of 4L to 6L α-Te, but they are still comparable with that of monolayer PtSe$_2$ ($10^3$–$10^4$ cm$^2$/Vs) [27] and appreciably larger than those of few-layer BP ($10^2$–$10^3$ cm$^2$/Vs). The mobilities along the $z$ direction are generally smaller than those along the $y$ direction. It could be inferred that the carrier mobility shall continue increasing with respect to the enlarged layer thickness according to the formula of carrier mobility and our results from 2L to 6L. Here, we only consider the scattering effect from the long wavelength acoustic phonons which is an ideal case. The experimental mobility is, however, affected by other aspects like gating efficiency, impurity scattering, charge traps, contacts and among the others. Given the present optimal thickness of ~15 nm, as revealed experimentally [58], it would be interesting to deduce an optimal thickness in further optimized devices.

Table 1 Predicted phonon-limited carrier mobility. Here, carrier types "e" and "h" denote "electron" and "hole", respectively. Variable $n_L$ represents the number of layers, $m_y^*$ and $m_z^*$ are carrier effective masses for directions $y$ and $z$, respectively, $E_{1y}$ ($E_{1z}$) and $C_y$ ($C_z$) are the deformation potential and elastic modulus for the $y$ ($z$) direction. Mobilities $\mu_y$ and $\mu_z$ were calculated using Eq. (1) with the temperature $T$ set to 300 K.

| Carrier type | $n_L$ | $m_y^*$ ($m_0$) | $m_z^*$ ($m_0$) | $E_{1y}$ (eV) | $E_{1z}$ (eV) | $C_y$ (GPa) | $C_z$ (GPa) | $\mu_y$ ($10^3$ cm$^2$/Vs) | $\mu_z$ ($10^3$ cm$^2$/Vs) |
|---|---|---|---|---|---|---|---|---|---|
| h | 2 | 0.58 | 0.36 | −0.57±0.03 | −1.28±0.02 | 26.2 | 32.5 | 3.71–4.62 | 2.50–2.70 |
|   | 3 | 0.40 | 0.38 | −0.32±0.02 | −2.22±0.02 | 24.6 | 36.3 | 31.6–41.7 | 1.23–1.27 |
|   | 4 | 0.36 | 0.34 | −[1] | −2.29±0.04 | 24.1 | 37.9 | – | 1.97–2.11 |
|   | 5 | 0.40 | 0.33 | 0.20±0.08 | −2.12±0.02 | 23.3 | 38.8 | 70.9–331 | 2.95–3.03 |
|   | 6 | 0.30 | 0.25 | 0.12±0.06 | −2.14±0.06 | 22.9 | 39.4 | 341–2690 | 5.78–6.44 |
| e | 2 | 0.85 | 0.87 | −0.93±0.01 | 0.25±0.01 | 26.2 | 32.5 | 0.68–0.73 | 10.8–13.0 |
|   | 3 | 0.49 | 1.68 | −0.84±0.03 | 0.94±0.05 | 24.6 | 36.3 | 3.30–3.85 | 0.29–0.35 |
|   | 4 | 0.23 | 0.18 | 1.52±0.03 | −6.26±0.03 | 24.1 | 37.9 | 6.37–6.94 | 0.96–0.98 |
|   | 5 | 0.28 | 0.15 | 1.30±0.03 | −6.70±0.02 | 23.3 | 38.8 | 6.40–7.00 | 1.62–1.64 |
|   | 6 | 0.24 | 0.13 | 1.54±0.03 | −6.79±0.03 | 22.9 | 39.4 | 7.38–8.06 | 2.50–2.55 |

[1] The 4L appears a transition thickness that the sign reverses for the deformation potential of the VB along y. The fitting error was, therefore, too large to reliably deduce $E_{1y}$ for hole, therefore, we left this value and the associated hole mobility blank.

### 3.5. Broadband and strong light absorption

Although the in-plane geometric anisotropy leads to significant direction-dependent behaviors for several physical properties, the anisotropy of optical absorption, however, eliminates for 2L-α-Te and slightly restores at 6L, as shown in Fig. 4a and 4b. The near isotropy of light absorption suggests FL-α-Te a potentially good light absorber. Our calculated results show that the absorbance is rather high for the normal incident light linearly polarized along the two in-plane directions (*y* and *z*), as red solid line and blue dashed line shown in Fig. 4a and 4b, roughly 2%–3% per layer at 1.6 eV and 6%–9% at 3.2 eV, which are nearly twice to three-times to that of BP [4]. This extraordinarily high light-absorbance suggests FL-α-Te of great potential for optical applications. Fig. 4c shows the absorbance of the 2L to 6L and bulk α-Te for the light linearly polarized along the *y* (non-covalent) direction. It shows that the band-edge for light absorption is roughly 0.2 eV larger than the indirect bandgap. This difference suggests the inter-band transition and carrier transfer of the excited FL-α-Te may occur, which are of great interest to be investigated in detail. A pronounced absorption peak around 1.3 to 1.6 eV is observable in 2L-α-Te (Fig. 4a), which corresponds to the absorption edge of directly inter-band transitions between the valence and conduction bands. Since the interlayer electronic hybridization and thus the electronic band dispersion are rather weak in 2L (Fig. 2d), the lowest transition energies are mostly around 1.3 to 1.6 eV at different *k* points in the reciprocal space, leading to speedily increased absorbance. This absorption edge is apprecibly smoothed in 3L and thicker layers owing to larger electronic band dispersion induced by stronger interlayer electronic hybridization (Fig. S4 online). As a result, the absorbance is layer-dependent that the absorption efficiency substantially increases with the decreased thickness (Fig. 4c and Fig. S8 online). For instance, the absorbance per layer enhance from 4.4% (bulk) to 7.3% (2L) for the green light (2.42 eV) and 4.3% (bulk) to 8.4% (2L) for the violet light (3.24 eV) with the polarization direction along *y*, a non-covalently bonded direction; this is another piece of evidence for the strong interlayer and inter-chain couplings in FL-α-Te. The layer- and direction-dependent light absorption also suggest the FL-α-Te is the nanostructured form holding the highest light absorption efficiency, which is even superior to the FL-α'-Te nanoplate.

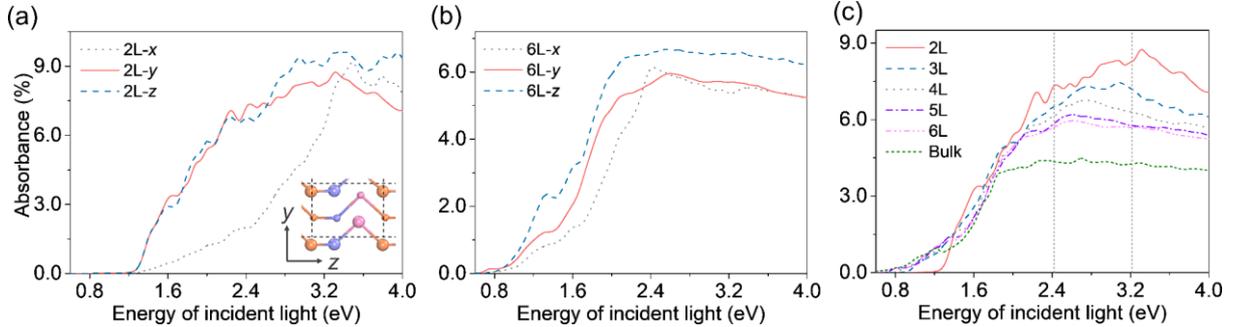

Fig. 4. (Color online) Optical absorption spectra of FL-α-Te. (a), (b) Absorbance per layer of 2L- and 6L-α-Te with the polarization direction of incident light along *x*, *y* and *z*, respectively. (c) Absorbance per layer with the polarization direction of incident light along *y* for FL-α-Te with the thickness varying from 2L to 6L and bulk.

### 3.6. Mechanical properties and vibrational frequencies

Helically appeared Te atoms in FL-α-Te imply that the ideal strength of it should be significant in comparison with previously known 2D materials, even BP few-layers. The ideal strength along the *z* direction is up to 18% (based on strain) – 26% (based on energy) for 2L- to 4L-α-Te, and the value of bulk is up to 38% (based on energy), as shown in Fig. S9 (online). Beyond that strain, the helical chain transforms to a new phase, namely the δ phase with space group $P2_1/m$ (No. 11) for 2L, which will be discussed in detail elsewhere. The bulk form can sustain compressive strain up to 14% that larger compressive strain breaks Te-Te bonds. For few-layers, however, the β-phase emerges if the compressive strain is up to 10%–12%.

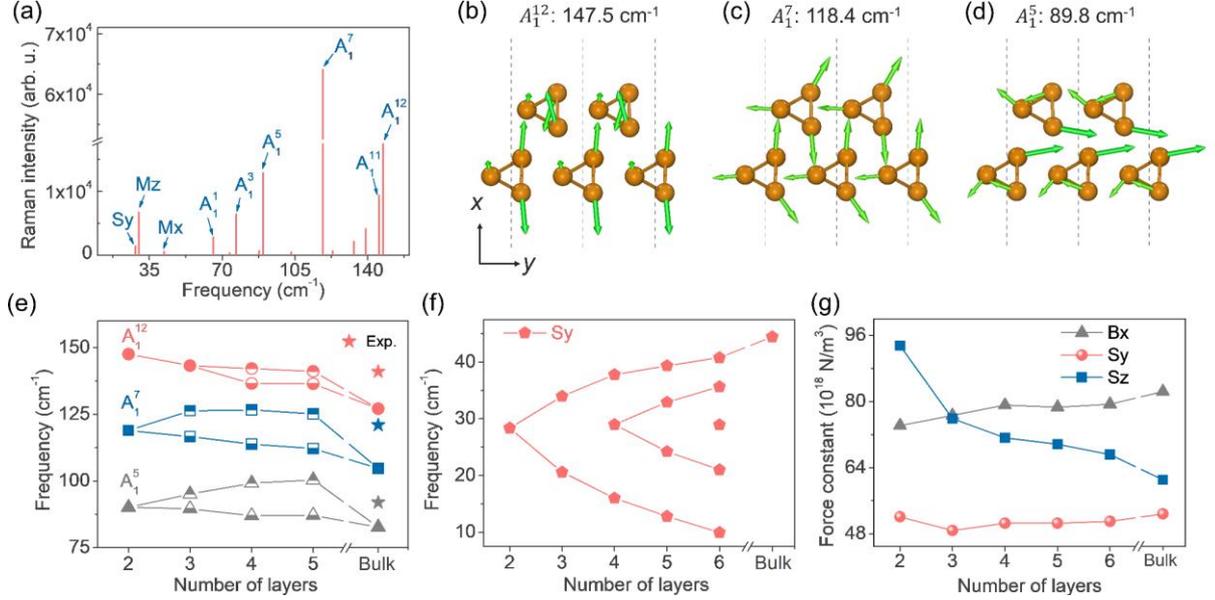

Fig. 5 (Color online) Vibrational properties of FL-α-Te. (a) Raman intensity of 2L-α-Te. (b)–(d) Schematic diagrams of vibrational displacements for modes $A_1^{12}$, $A_1^7$ and $A_1^5$ of 2L-α-Te. (e) The $A_1^{12}$, $A_1^7$ and $A_1^5$ modes show a frequency-splitting forming two branches and anomalous blue- or red-shift of few-layer α-Te. (f) Layer-dependence of vibration frequencies at the G point for the $S_y$ mode. (g) Layer-dependent force constants for shear and breathing modes of α-Te.

Raman spectroscopy is of vital importance for characterization of layered materials. Fig. 5a shows the theoretical Raman intensity of 2L-α-Te as a function of vibrational frequency. The symmetries of all modes are rather low, and each optical mode is represented by $A_1^x$. There are three most pronounced optical modes showing strong Raman intensity, namely $A_1^5$ (Fig. 5b), $A_1^7$ (Fig. 5c) and $A_1^{12}$ (Fig. 5d). They are stemmed from $E_1^1$, $A_1$ and $E_1^2$ modes of bulk Te, respectively (Fig. S10 online). Mode $A_1^{11}$ (Fig. S11b online), with the fourth strongest Raman intensity, is a pair mode of mode $A_1^{12}$. Fig. 5b–5d illustrate the vibrational displacements of the three Raman-activated optical (RAO) modes, which are highly comparable with those of their bulk forms. Fig. 5e shows the frequencies of these RAO modes as a function of layer thickness. For few-layers thicker than 2L, the frequencies of these modes substantially change and develop into two branches, namely a lower (symbols with lower-half filled) and a higher (symbols with upper-half filled) branches. In particular, the two branches of modes $A_1^{12}$ and $A_1^7$ and the lower branch of mode $A_1^5$ exhibit redshift from 2L to 6L while the higher branch of mode $A_1^5$ shows blueshift. The averaged value of each mode is comparable with the experimentally measured values. Particularly, theoretically predicted values of 2L-α-Te are 89.8, 118.4 and 147.5 cm$^{-1}$, respectively, within 4 cm$^{-1}$ difference from the measured values of 94, 122 and 144 cm$^{-1}$ [59]. If we compare the frequencies of 2L with those of the bulk form, the all three modes show significant redshifts, in different from conventional weakly coupled 2D materials where the frequencies of optical phonons are usually lowered in few-layers [60]. The shift is up to 20 cm$^{-1}$ from 2L to bulk, which is exceptional even in comparison with the BP value of 3 cm$^{-1}$, suggesting even stronger interlayer coupling in FL-Te than in FL-BP.

These splitting and anomalous shifts of frequency are relevant with the distorted of vibration motions of inner layers. The vibrational displacements of these three RAO modes may be distorted at inner layers, especially for thicker layers (Fig. S12, S13 online). For modes and $A_1^5$, if the displacements of the inner layers change to represent a stiffer motion, the distorted mode thus becomes the higher branch of the corresponding RAO mode. However, if these layers keep the original motions, the frequency is thus lowered with respect to layer thickness since collectively motion with additional layers reduces the vibrational frequency. The lowered frequencies thus give rise in the lower branches of the associated RAO modes. Mode $A_1^{12}$, the highest frequency RAO mode, is always distorted to mix with softer modes at inner layers. Therefore, its frequencies at always lowered in thicker layers. The detailed reasons for the red- and blue-shifts in the few-layers are rather complicated and are beyond the scope of this work.

In brief, these anomalous frequency shifts are the results of the competition between surface effects and strong interlayer coupling, similar to those found in BP [11]. These results are quite unique even compared with those found in BP, a 2D material having anomalous vibrational properties. In light of these, the substantial layer-dependent frequency of RAO modes could be employed a descriptor to identify the number of layers as previous adopted in graphene layers [61].

It is a fingerprint feature for the interlayer acoustic (IA) vibrations of FL-α-Te that the appearance of shear-breathing mixed modes (Fig. S14 online). There are three low frequency IA modes in the bilayer. Their frequencies are 28.3 cm$^{-1}$ ($S_y$, shear mode along $y$), 29.9 cm$^{-1}$ ($M_x$, shear-breathing mixed mode primarily along $x$) and 42.1 cm$^{-1}$ ($M_z$, shear-breathing mixed mode primarily along $z$). The $S_y$ mode is Raman-activated and its layer dependence mostly follows the Davydov splitting rule, as shown in Fig. 5f. Interlayer force constants projected on the $x$, $y$ and $z$ directions are thus deduced from the dynamic matrixes constructed by these low-frequency IA modes. It is remarkable that the differences among the force constants of the three directions are smaller than most of other layered materials. The force constant along $y$ is nearly thickness independent around $50 \times 10^{18}$ N/m$^3$, while those for the other two directions are correlated together and vary between $(60–90) \times 10^{18}$ N/m$^3$ depending on the layer thickness, as shown in Fig. 5g. A crossover of the force constants projected for the interlayer shear and breathing modes was found at 3L, owing to inter-chain covalent-like quasi-bonding and the competition between surface and inner layers. This crossover indicates that the rigidity of the glide motion along $z$ is even higher than that of the breathing motion along $x$ in thinner layers (<4L), indicating unexpected flexibility of 2L and 3L. The force constant for the explicit shear mode ($S_y$) is moderate, comparable with that of PtS$_2$ [25]. However, the other shear mode along Te chains (the $z$ direction) was so strong, especially in thinner layers, that correlates with the breathing mode introducing the two shear-breathing mixed modes.

### 3.7. Environmental stability of few-layer alpha-Te

Environmental stability is a key issue for elementary semiconductors, e.g. BP [48, 62]. Here, we considered the interaction between FL-α-Te and $O_2$ or $H_2O$ molecule. Fig. 6a shows the most stable configuration of $O_2$ physisorbed on an α-Te bilayer with the adsorption energy of – 0.22 eV. This configuration is at least 38 meV more stable than other configurations (see Fig. S15 online). An energy release of 2.09 eV was found after of $O_2$ dissociated into two chemisorbed O atoms on this bilayer, as shown in Fig. 6e. This configuration is the most stable chemisorption configuration among 23 considered configurations, which cover all possible single O adsorption sites (Fig. S16 and S17 online). The secondary stable chemisorbed configuration was shown in Fig. 6f, which is 0.28 eV less stable. Their dissociative reaction pathways were illustrated in Fig. 6a, 6c, 6d, 6g, 6h, with the configurations of transition states and the barrier energy plotted (Fig. 6g and 6h). It turns out the barrier is at least 0.85 eV and up to 0.94 eV, which is sufficient to prevent α-Te being oxidized at ambient conditions. However, if the temperature increases to over 500 K, the oxidization process becomes appreciable. Light induced hot electrons may also trigger the oxidization at room temperature, which will be discussed elsewhere. Fig. 6b shows the most stable configuration of $H_2O$ physisorbed on α-Te. Its adsorption energy is 0.29 eV, several meV larger than that on BP and the intermolecular interaction energy of $H_2O$ [48], which results in α-Te slightly hydrophilic with a contact angle of 76°. The splitting of water is an endothermic reaction with an energy loss of 1.36 eV.

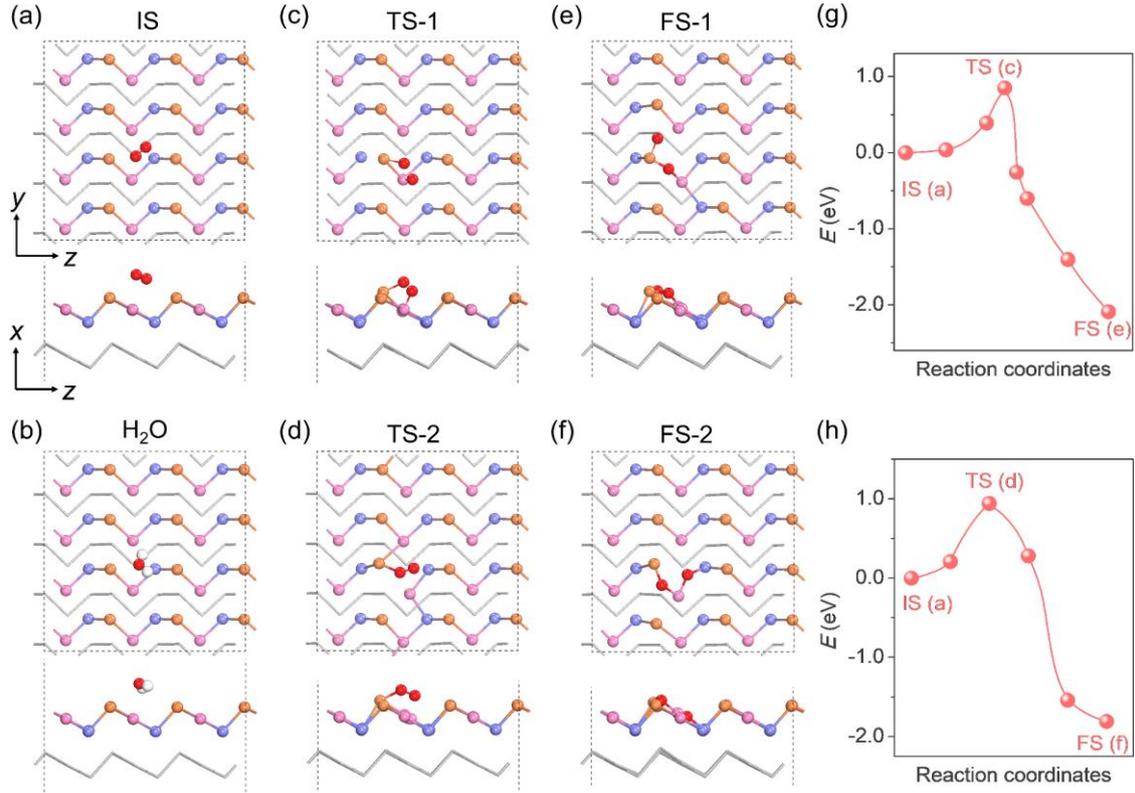

Fig. 6 (Color online) Environmental stability of FL-α-Te. (a), (b) Atomic structures of physisorbed $O_2$ (a) and $H_2O$ (b) on bilayer α-Te. (c)–(f) The configurations of transition states ((c), (d)) and final states ((e), (f)) for oxidation pathways I and II, respectively. (g), (h) The energy profile of reaction pathways I (g) and II (h) for $O_2$ on α-Te.

## 4. Conclusion

In summary, we carried out density functional theory calculations to predict physical properties of FL-α-Te which was found the most stable in 2L and thicker layers among other layered phases. It is, to the best of our knowledge, the first comprehensively investigated 1D-like, in-plane vdW- (CLQB-) bonded and solution-based-synthesizable layered material. It succeeds and/or is superior to most of striking properties of BP, e.g. tunable moderate bandgap, intrinsic anisotropy, extraordinary hole mobility, strongly optical absorption, highly stretchable and nearly direct bandgap. The high mobility and optical absorption may imply its potential application for sensitive photo-detector. Anisotropic, complex and layer-dependent electron and hole pockets were found for FL-α-Te, suggesting more suitable electronic structures of Te few-layers than that of Te nanowires for potentially high thermo-electric performance. In addition, the nearly isotropic in-plane optical absorbance yields stronger absorption than that of those nanowires or nanoplates in the α'-phase, which was a result of the inter-chain (intra-layer) CLQB as extended from previously found interlayer cases of other 2D layers. The highest mobility was found in the CLQB dominant direction; this is largely out of our exception and conceptually revises our previous understanding for weakly bound systems. The CLQB additionally results in shear-breathing mixed vibrational modes and thus a crossover of the interlayer force constants of shear and breathing modes. This CLQB also suggests much easier inter-chain or inter-layer movement (immigration) of Te atoms due to its comparable strength with and similar covalent characteristic to the intra-chain Te-Te covalent bonding. The large oxidization barrier, indicating good environmental stability, is superior to the substantial drawback of BP. Since the α phase is chiral and has two polarization directions, these results imply FL-Te may have piezoelectric effect due to strain induced α-β transitions, ferroelectric effect led by transitions between two chiral directions of α-Te through β-Te and electrostrictive phenomena because of charge induced α-β transitions. All these striking results demonstrate FL-α-Te a highly promising elementary semiconductor for diverse applications and drive us to explore the unknown fields of 1D-like layered materials.

**Note added**: During the preparation for this manuscript, we became aware that few-layer α-Te samples have been synthesized in large scale and fabricated into electronic devices [58] and the synthesis of it was also realized by molecular beam epitaxy [63, 64], as well as the β- and γ-phases were theoretically discussed [65].

## Conflict of interest
The authors declare that they have no conflict of interest.

## Acknowledgments

This work was supported by the National Natural Science Foundation of China (11274380, 91433103, 11622437, 61674171, and 61761166009), the Fundamental Research Funds for the Central Universities of China and the Research Funds of Renmin University of China (16XNLQ01), and The Hong Kong Polytechnic University (G-SB53). J.Q. and C.W. were supported by the Outstanding Innovative Talents Cultivation Funded Programs 2016 and 2017 of Renmin University of China, respectively. Calculations were performed at the Physics Lab of High-Performance Computing of Renmin University of China.

# Supplementary Materials: Strong inter-chain coupling and extraordinarily high carrier mobility in stretchable Te few-layers

Jingsi Qiao[1,2], Yuhao Pan[1], Feng Yang[1], Cong Wang[1], Yang Chai[2] and Wei Ji[1, *]

[1]*Beijing Key Laboratory of Optoelectronic Functional Materials & Micro-Nano Devices, Department of Physics, Renmin University of China, Beijing 100872, P. R. China*

[2]*Department of Applied Physics, The Hong Kong Polytechnic University, Hung Hom, Kowloon, Hong Kong, P. R. China*

* wji@ruc.edu.cn

Including Fig. S1-S21 and Table S1-S4:

**Table S1** provides the lattice constants of few-layer and bulk α-Te calculated using the optB88-vdW, optB86b-vdW, optPBE-vdW, SCAN and SCAN-rVV10 functionals.

**Table S2** shows the predicted phonon-limited carrier mobility using another model.

**Table S3** provides the bonding energies of $O_2$ and adsorption energies of physisorption and chemisorption configurations calculated using the optB88-vdW, SCAN and SCAN-rVV10 functionals.

**Fig. S1** illustrates of structural phase transitions among different Te few-layer allropes.

**Fig. S2** shows the phase diagram of 2L-α-, -β- and -γ-Te and their transitions under in-plane biaxial strain or uniaxial strain along the *z* and *y* directions.

**Fig. S3** provides the geometric and electronic structures of one-dimensional Te chains and clusters.

**Fig. S4** illustrates the supplementary bandstructures for FL-α-Te.

**Table S4** provides the exact VBM and CBM positions in the Brillouin zones of FL- and bulk-α-Te.

**Fig. S5-S6** illustrate the supplementary schematic drawings of energy surface and visualized wavefunctions for FL-α-Te.

**Supplementary Note 1** describes a one-dimensional quantum well model for deriving the relationship between the bandgaps and the number of layers.

**Fig. S7** shows the fitted curve of the bandgaps of α-Te as a function of the number of layers.

**Fig. S8** provides the supplementary optical absorption spectra of FL-α-Te.

**Fig. S9** shows the relative energy-strain response and ideal strengths of bulk- and FL-α-Te.

**Fig. S10-S14** illustrate the supplementary Raman activated modes in bulk-, 2L-, 3L- and 4L-α-Te.

**Fig. S15-S17** provide the representative configurations of physisorbed $O_2$ and chemisorbed single O and $O_2$ on 2L-α-Te.

**Fig. S18-S21** shows the relative errors in fitting the deformation potentials.

**Table S1.** Lattice constants of FL- and bulk-α-Te were calculated using the optB88-vdW, optB86b-vdW, optPBE-vdW, SCAN and SCAN-rVV10 functionals. Here, $b$ and $c$ are marked on Fig. 1a and 1b and $n_L$ represents the number of layers.

| $n_L$ | 2L | | 3L | | Bulk | |
|---|---|---|---|---|---|---|
| Lattice constant (Å) | $b$ | $c$ | $b$ | $c$ | $b$ | $c$ |
| optB88-vdW | 4.36 | 5.81 | 4.40 | 5.88 | 4.48 | 5.99 |
| optB86-vdW | 4.27 | 5.80 | 4.31 | 5.88 | 4.39 | 6.00 |
| optPBE-vdW | | | | | 4.55 | 6.00 |
| SCAN-rVV10 | 4.24 | 5.74 | 4.28 | 5.80 | 4.38 | 5.94 |
| SCAN | 4.31 | 5.79 | 4.36 | 5.84 | 4.46 | 5.93 |
| Exp.[a][ref. 49] | | | | | 4.46 | 5.92 |
| Exp.[b][ref. 50] | | | | | 4.46 | 5.93 |

| $n_L$ | 4L | | 5L | | 6L | |
|---|---|---|---|---|---|---|
| Lattice constant (Å) | $b$ | $c$ | $b$ | $c$ | $b$ | $c$ |
| optB88-vdW | 4.42 | 5.91 | 4.44 | 5.93 | 4.44 | 5.94 |
| SCAN-rVV10 | 4.30 | 5.84 | 4.31 | 5.85 | 4.32 | 5.86 |

**Table S2.** Predicted phonon-limited carrier mobility by $\mu_{2D} = \dfrac{e\hbar^3 C_{2D}}{k_B T m_e^* m_d (E_1^i)^2}$ (2), where $m_e^*$ is the effective mass along the transport direction and $m_d$ is the density-of-state mass determined by $m_d = \sqrt{m_y m_z}$. $C_{2D}$ is the elastic modulus of the longitudinal strain in the propagation direction, which derived by $(E - E_0)/S_0 = C(\Delta l/l_0)^2 /2$, where $E$ is the total energy and $S_0$ is the lattice volume at the equilibrium for 2D systems. Carrier types 'e' and 'h' denote 'electron' and 'hole', respectively. $n_L$ represents the number of layers, $m_y^*$ and $m_z^*$ are carrier effective masses for directions $y$ and $z$, respectively, $E_{1y}$ ($E_{1z}$) and $C_{y\_2D}$ ($C_{z\_2D}$) are the deformation potential and 2D elastic modulus for the $y$ ($z$) direction. Mobilities $\mu_y$ and $\mu_z$ were calculated using equation (2) with the temperature $T$ set to 300 K.
See $E_1$ and $^1$ in the method section and Table 1 of the manuscript.

| Carrier type | $n_L$ | $m_y^*$ | $m_z^*$ | $E_{1y}$ | $E_{1z}$ | $C_{y\_2D}$ | $C_{z\_2D}$ | $\mu_y$ | $\mu_z$ |
|---|---|---|---|---|---|---|---|---|---|
| | | ($m_0$) | | (eV) | | (J/m$^2$) | | (10$^3$ cm$^2$/Vs) | |
| h | 2 | 0.58 | 0.36 | -0.57±0.03 | -1.28±0.02 | 21.5 | 26.5 | 4.82~6.00 | 2.02~2.18 |
| | 3 | 0.40 | 0.38 | -0.32±0.02 | -2.22±0.02 | 29.5 | 43.4 | 33.9~44.8 | 1.24~1.28 |
| | 4 | 0.36 | 0.34 | -[1] | -2.29±0.04 | 38.1 | 59.8 | - | 2.03~2.17 |
| | 5 | 0.40 | 0.33 | 0.20±0.08 | -2.12±0.02 | 45.5 | 75.9 | 86.0~401 | 2.94~3.02 |
| | 6 | 0.30 | 0.25 | 0.12±0.06 | -2.14±0.06 | 53.6 | 92.1 | 414~3271 | 5.83~6.49 |
| e | 2 | 0.85 | 0.87 | -0.93±0.01 | 0.25±0.01 | 17.0 | 25.7 | 0.70~0.75 | 11.4~13.7 |
| | 3 | 0.49 | 1.68 | -0.84±0.03 | 0.94±0.05 | 23.8 | 42.8 | 1.87~2.19 | 0.63~0.77 |
| | 4 | 0.23 | 0.18 | 1.52±0.03 | -6.26±0.03 | 31.1 | 59.1 | 7.41~8.07 | 0.88~0.90 |
| | 5 | 0.28 | 0.15 | 1.30±0.03 | -6.70±0.02 | 37.4 | 75.1 | 9.59~10.49 | 1.21~1.23 |
| | 6 | 0.24 | 0.13 | 1.54±0.03 | -6.79±0.03 | 44.2 | 91.8 | 11.1~12.1 | 1.91~1.95 |

**Table S3.** Bonding energies of O$_2$ molecule calculated using the optB88-vdW, SCAN and SCAN-rVV10 functionals. The adsorption energies of a physisorption (P1, Fig 6a/ Fig. S9a) and two chemisorption (C1, Fig.6e/ Fig. 11a and C4, Fig. 6f/ Fig. S11d) configurations were calculated using the optB88-vdW and SCAN-rVV10 functionals, respectively. Our adsorption energies, revealed with the optB88-vdW functional, are comparable with the values unveiled by the SCAN and SCAN-rVV10 functionals, which suggests the reliability of our calculation. In terms of the physisorption state (initial adsorption), O$_2$ was found still in its triplet state holding the 2 $\mu_B$ magnetic moment.

| Bonding energy of O$_2$ | | |
|---|---|---|
| optB88-vdW | SCAN | SCAN-rVV10 |
| -6.00 | -5.59 | -5.60 |
| Adsorption energy (eV) | | |
| Configurations | optB88-vdW | SCAN-rVV10 |
| P1 | -0.22 | -0.18 |
| C1 | -2.31 | -2.39 |
| C4 | -2.03 | -2.09 |

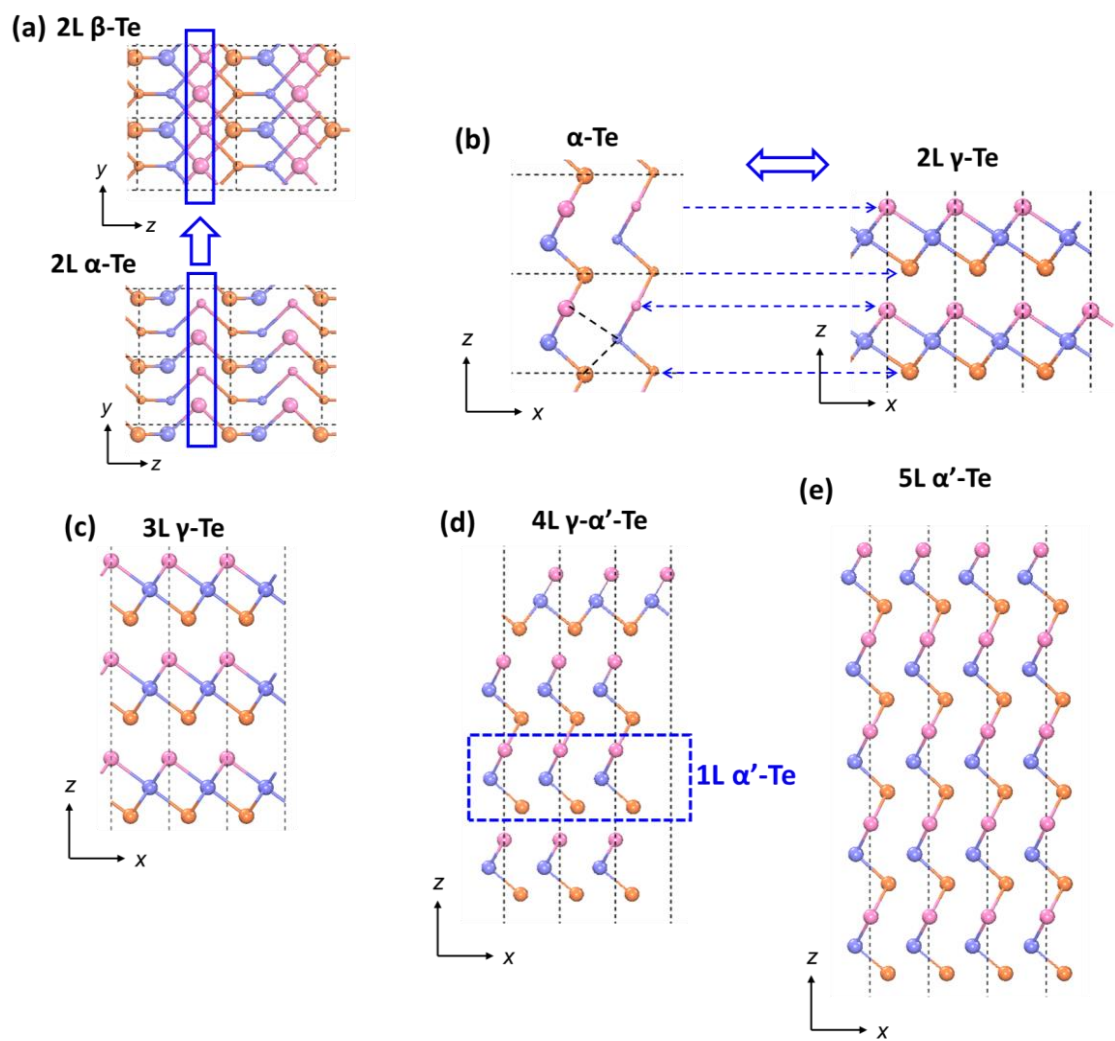

**Fig. S1.** Illustration of the structural phase transitions among different phases of Te few-layers. (a) A transition from the α- to β-phase is realized by moving those Te atoms marked with the blue rectangle along the +*y* direction. (b) Correspondence of Te atoms in the transition from the α- to γ-phase. (c) 3L γ-phase Te is more stable than the associated α'-phase. Here, we define three atomic layers as ``one layer'' of α'-phase as shown in the blue dashed rectangle in panel d. Few-layer α'-phase is formed by cutting the covalent bonds along the *z* direction of the Te chains in the α-phase bulk form. (d) fully relaxed 4L γ/α'-phase Te shows a γ-α' mixed configuration. (e) Fully relaxed 5L γ/α'-phase indicates α'-phase is more stable for 5L and thicker layers.

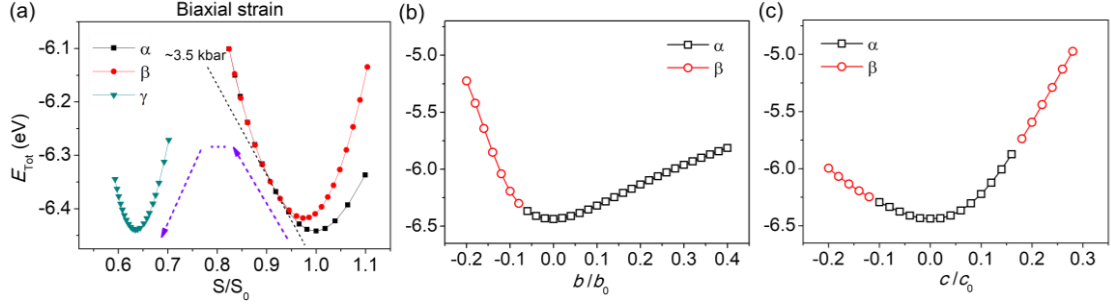

**Fig. S2.** (a) Phase diagram of 2L α-, β- and γ-Te as a function of in-plane area (S). $S_0$ is the area of 2L-α-Te in equilibrium. A phase transition from α- to β-Te occurs under an applied biaxial stress of 3.5 kbar, equivalent 7.5% area reduction, which is barrier-less. In addition, an α to γ phase transition can be achieved at a reasonable compressive strain. This transition is associated with a 63% collapse of the in-plane area and a fairly large lattice distortion. (b-c) Phase transitions between 2L-α- and -β-Te under uniaxial strain along the *y* (b) and *z* (c) direction.

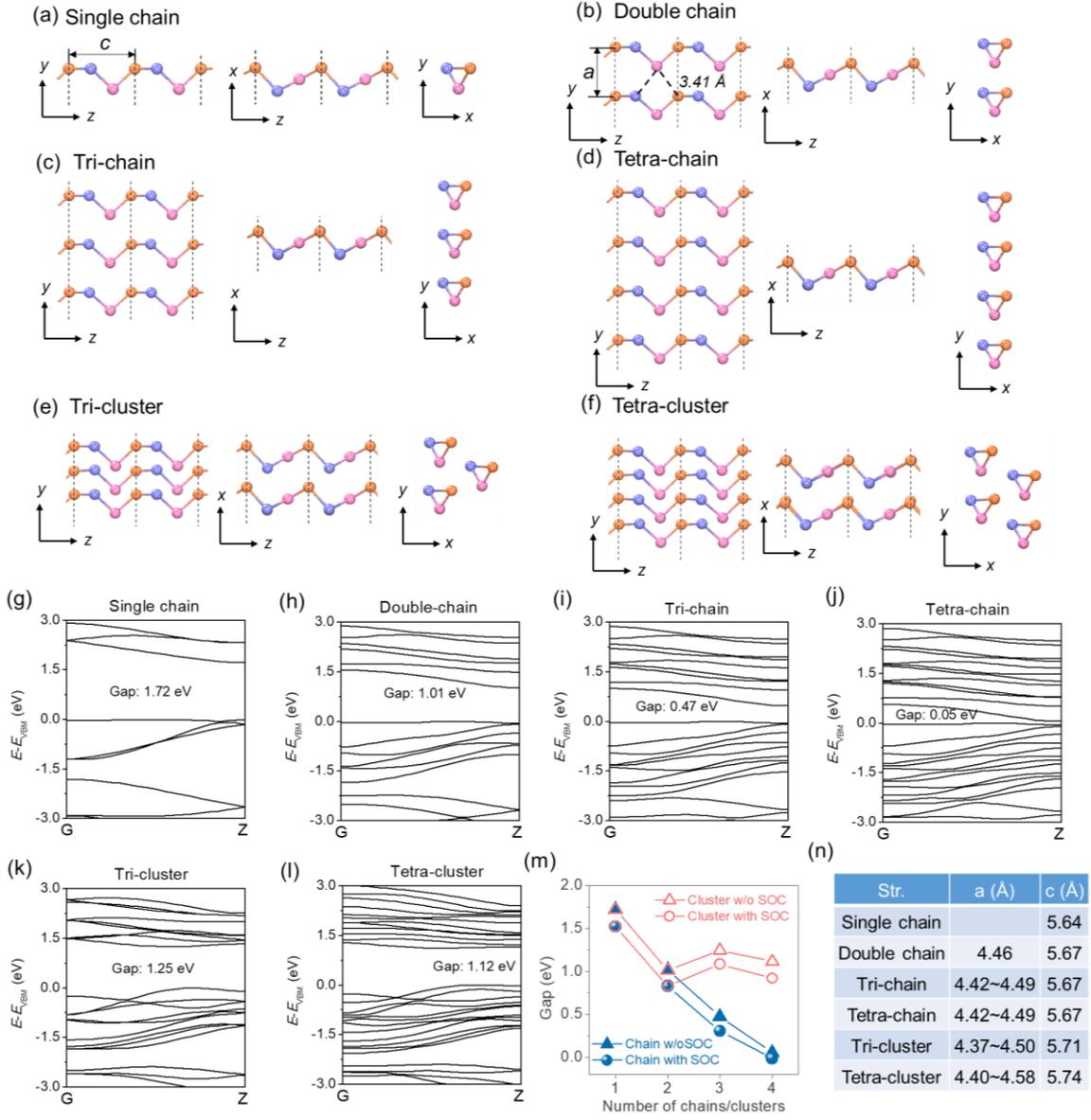

**Fig. S3.** Geometric and electronic structures of one-dimensional (1D) Te chains and clusters. (a-f) Crystal structures of single- (a), double- (b), tri- (c), tetra- (d) chains and tri- (e), tetra- (f) clusters. (g-l) Bandstructures of single- (g), double- (h), tri- (i), tetra- (j) chains and tri- (k), tetra- (l) clusters with 1.72, 1.01, 0.47, 0.05, 1.25, 1.12 eV bandgap, respectively. (m) Band evolution with the number of chains or clusters. The bandgap decreases with the increasing number of chains. (n) Lattice constants $a$ [marked in panel (b)] and $c$ [marked in panel (a)].

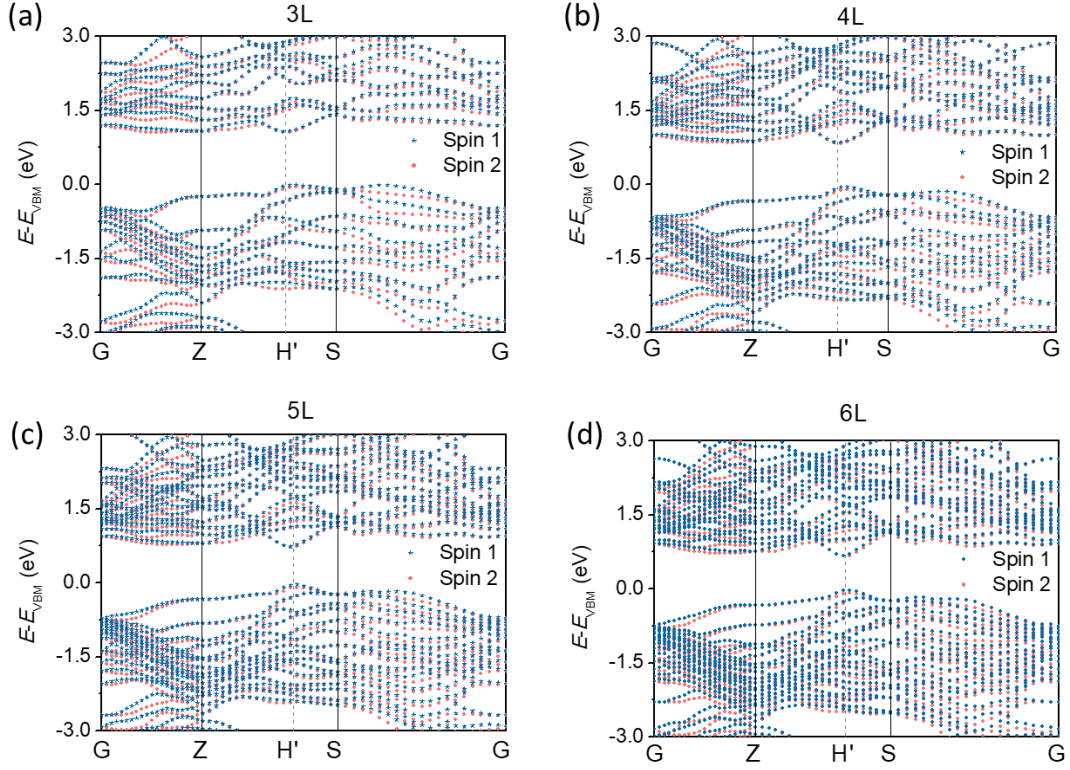

**Fig. S4.** Bandstructures of few-layer α-Te calculated using the HSE06 functional with inclusion of spin-orbit coupling (SOC). The Brillouin zone path is shown in Fig. 2b.

**Table S4.** Exact VBM and CBM positions in FL- and bulk-α-Te calculated using the HSE06 functional with the inclusion of SOC. Here, $n_L$ represents the number of layers.

| $n_L$ | VBM position | CBM position |
|---|---|---|
| 2 | (0.000  0.267  0.344) | (0.000  0.000  0.270) |
| 3 | (0.000  0.300  0.380) | (0.000  0.317  0.500) |
| 4 | (0.000  0.317  0.415) | (0.000  0.324  0.500) |
| 5 | (0.000  0.333  0.438) | (0.000  0.325  0.500) |
| 6 | (0.000  0.333  0.445) | (0.000  0.333  0.500) |
| Bulk | (0.333  0.333  0.485) | (0.333  0.333  0.495) |

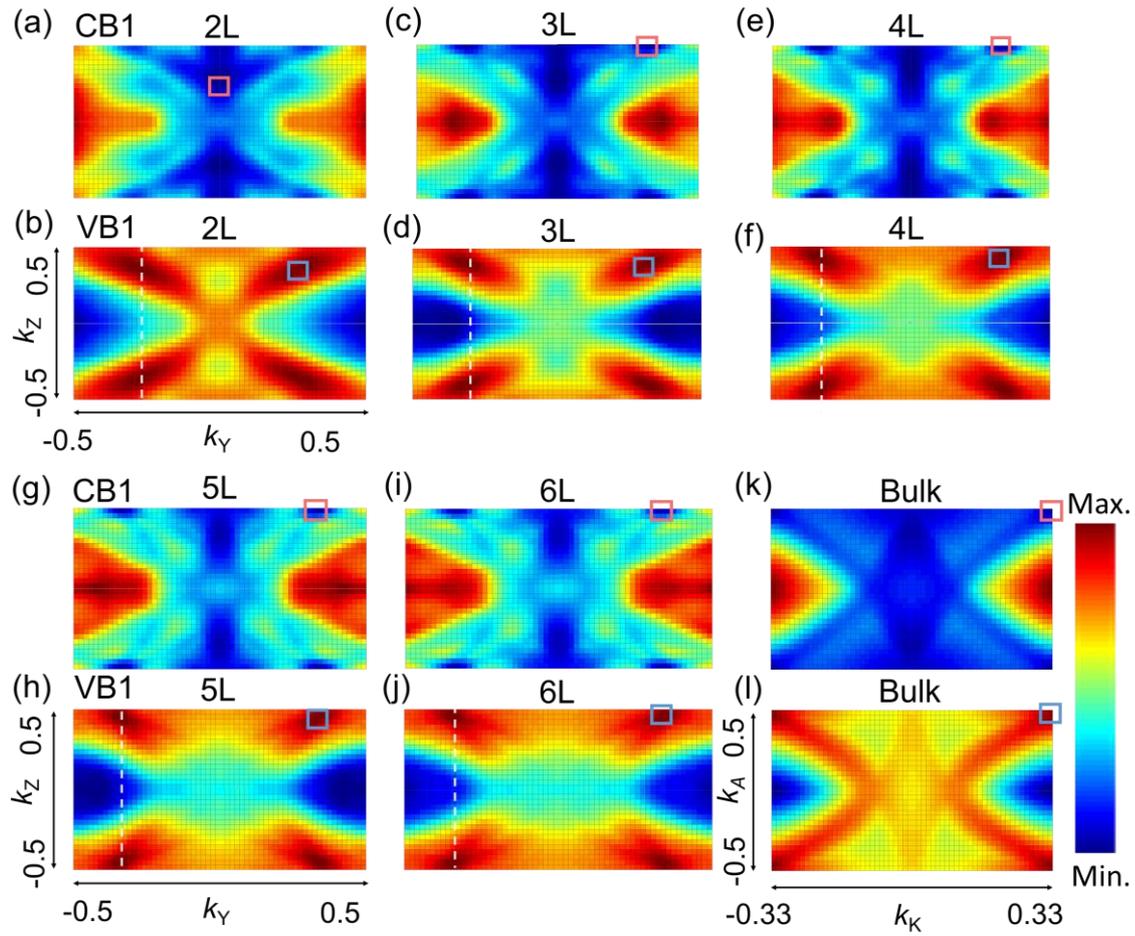

**Fig. S5.** Schematic drawings of energy surfaces for the lowest conduction band (CB1) and the highest valence band (VB1) from 2L (a, b) to bulk (k, l) along the *y-z* plane in the first Brillouin zone. Schematic data were derived from the results calculated using the optB88-vdW functional with the inclusion of SOC, which shares same tendency with the HSE06 functional.

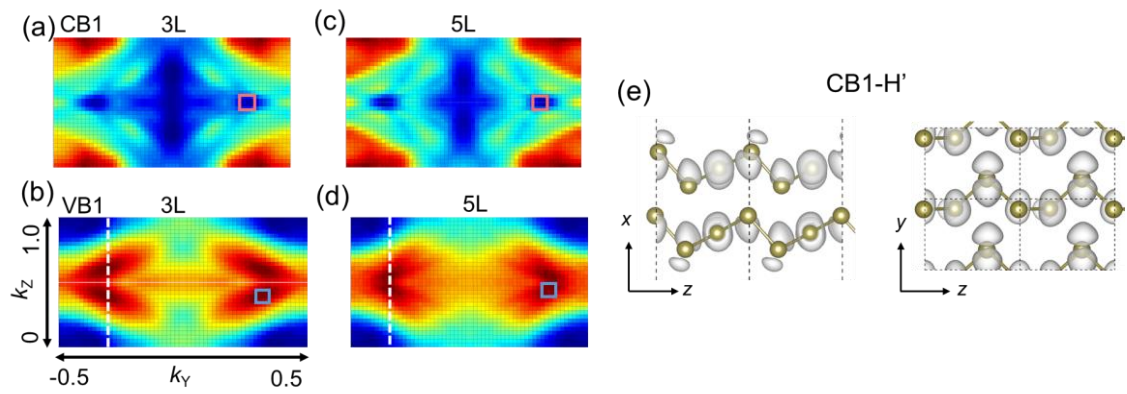

**Fig. S6.** (a-d) Schematic drawings of energy surface for the lowest conduction band (CB1) and the highest valence band (VB1) in 3L (a,b) and 5L (c,d) along the *y-z* plane in the reciprocal space. Schematic data were derived from the results calculated using the optB88-vdW functional with the inclusion of SOC, which shares same tendency with the HSE06 functional. (e) Visualized wavefunctions of CB1 (labeled in Fig. 2d) in 2L-α-Te (Fig. 2c) along the *x-z* and *x-y* planes with an isosurface of 0.0025 $e$ Bhor$^{-3}$.

**Supplementary Note1:**
The layer-dependent bandgap is primarily a result of the quantum confinement along the normal-plane direction in strongly interlayer coupled materials likely BP and Te. Here, we used a one-dimensional quantum well model as illustrated in ref. 56 [Surface Science Reports 39 (2000) 181-235].

The energy levels are represented by

$$E = \frac{\hbar^2 k^2}{2m}, k = \frac{n\pi}{d} (n = 1,2,3 \ldots) \ (1).$$

Here, $m$ is the free electron mass, $k$ is wave vectors, $d$ is the thickness of 2D materials. Thus, the energy of bandgap is

$$E_{gap} = E_{CBM} - E_{VBM} = \frac{\hbar^2 \pi^2}{2md^2}(2N+1) \ (2).$$

Here, $N$ is the quantum number of the highest valence band, which can be deduced by

$$N = N_{1L} n_L \ (4).$$

While, $N_{1L}$ and $n_L$ are the quantum number of the highest valence band in monolayer material and the number of layers, respectively.

Then, we assumed the thickness of few-layer 2D materials $d$ is

$$d \approx l_{1L} n_L \ (3).$$

Therefore, the bandgap of few-layer materials can be derived by

$$E_{gap} \approx \frac{\hbar^2 \pi^2}{2m l_{1L}^2}\left(\frac{2N_{1L}}{n_L} + \frac{1}{n_L^2}\right) \approx \frac{\hbar^2 \pi^2 N_{1L}}{m l_{1L}^2} \frac{1}{n_L} \propto \frac{1}{n_L} \ (5)$$

We found the bandgaps in few-layer materials are approximately direct proportional to the reciprocal of the number of layer, $1/n_L$. We show the fitted curve of the bandgaps of α-Te as a function of the number of layers in Fig. S7. The fitted slope is larger than the slope of the line connecting the data of 2L and 3L, but smaller than those for 4L-5L and 5L-6L. This deviation, we believe, is resulted from the layer-dependent lattice expansion which changes the value of $N_{1L}$. In light of this, the interlayer wavefunction overlap and the associated quantum confinement, together with the layer-dependent lattice expansion, result in the observed nearly reciprocal decay of the bandgap as a function of the layer-thickness.

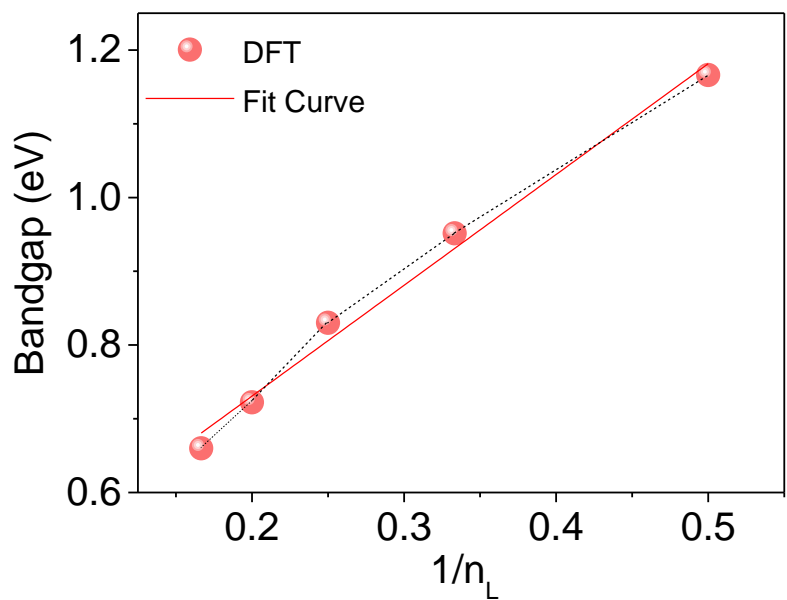

**Fig. S7** (a) Fitted curve of the bandgaps of FL-α-Te as a function of the number of layers. Here, $n_L$ is the number of layers.

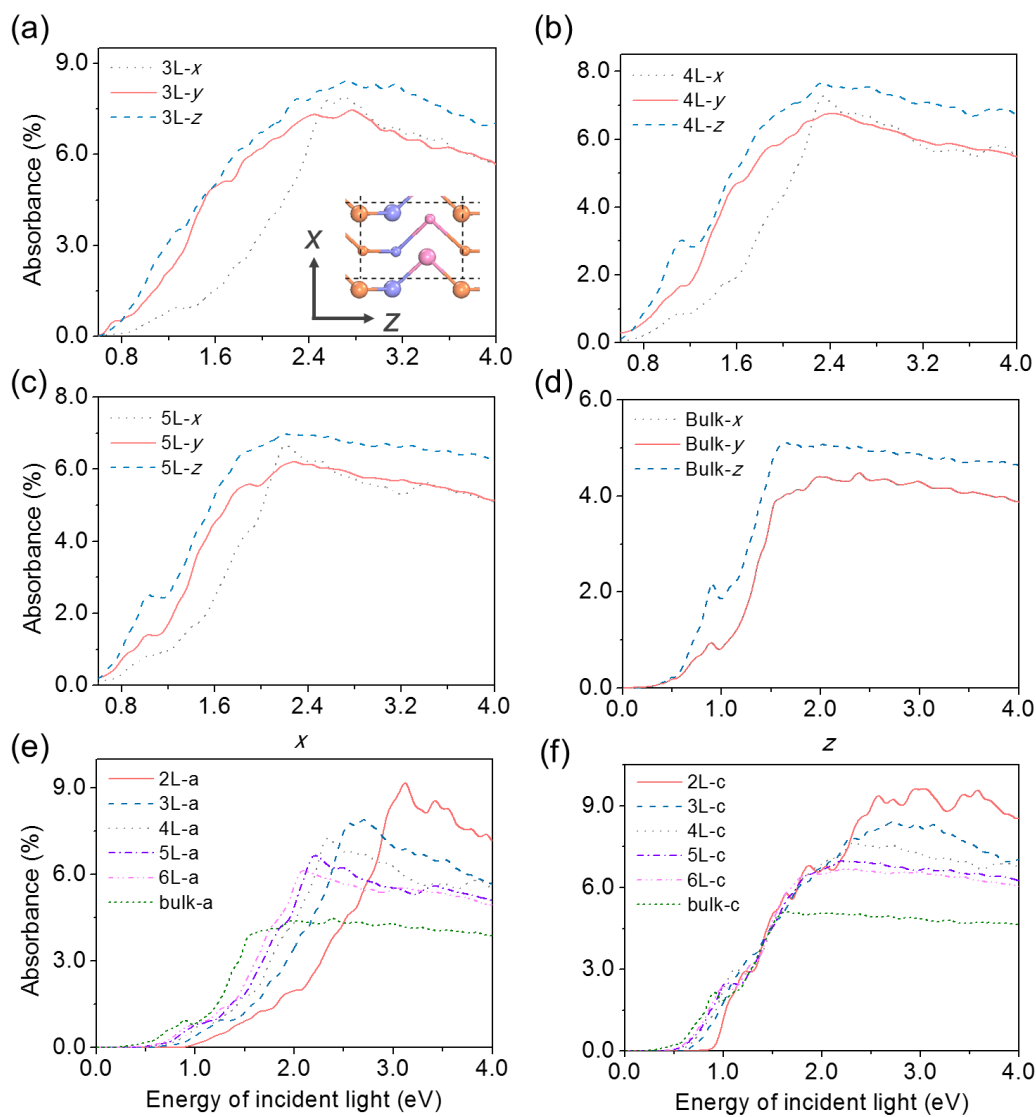

**Fig. S8.** (a-d) Absorbance per layer with the polarization direction of incident light along the $x$, $y$ and $z$ directions for 3L, 4L, 5L and bulk α-Te. (e-f) Absorbance per layer with the polarization direction of incident light along the $x$ and $z$ directions for α-Te with the thickness varying from 2L to bulk. Here, the energy of incident light of horizontal axis is the origin results calculated using the optB88-vdW functional without any shift of energy.

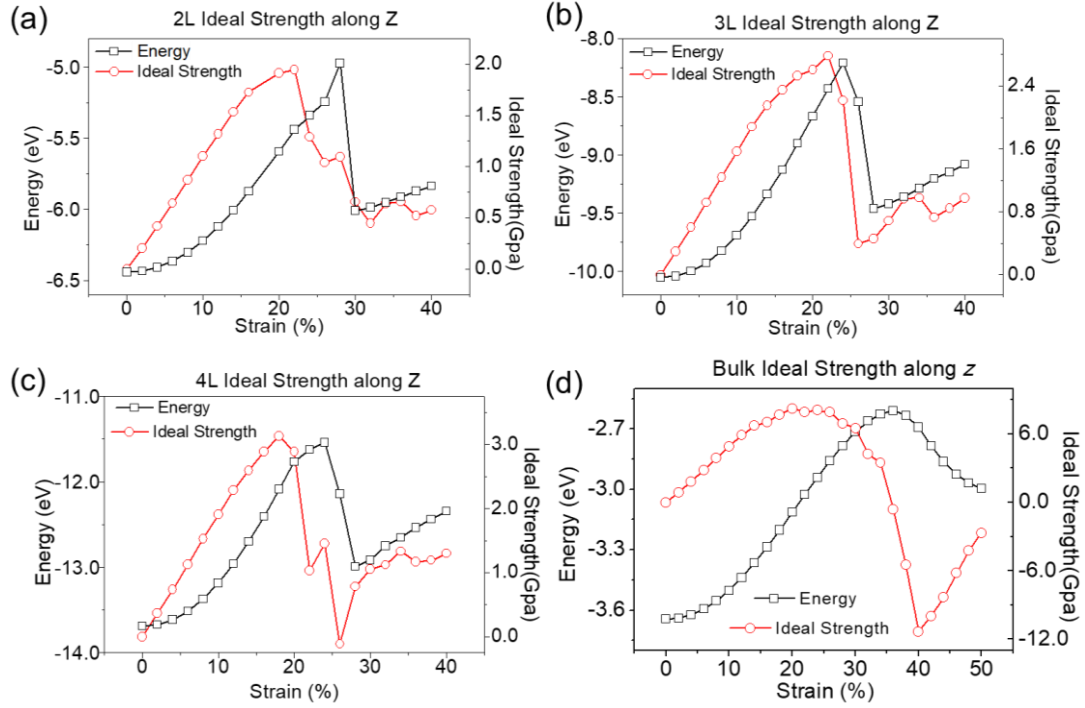

**Fig. S9.** Relative energy-strain response of bulk (a) and 2L-4L (b)-(d) α-Te along the *z* direction. Ideal strengths of the few-layers (2L-4L) are up to 28% and 44% for the bulk form.

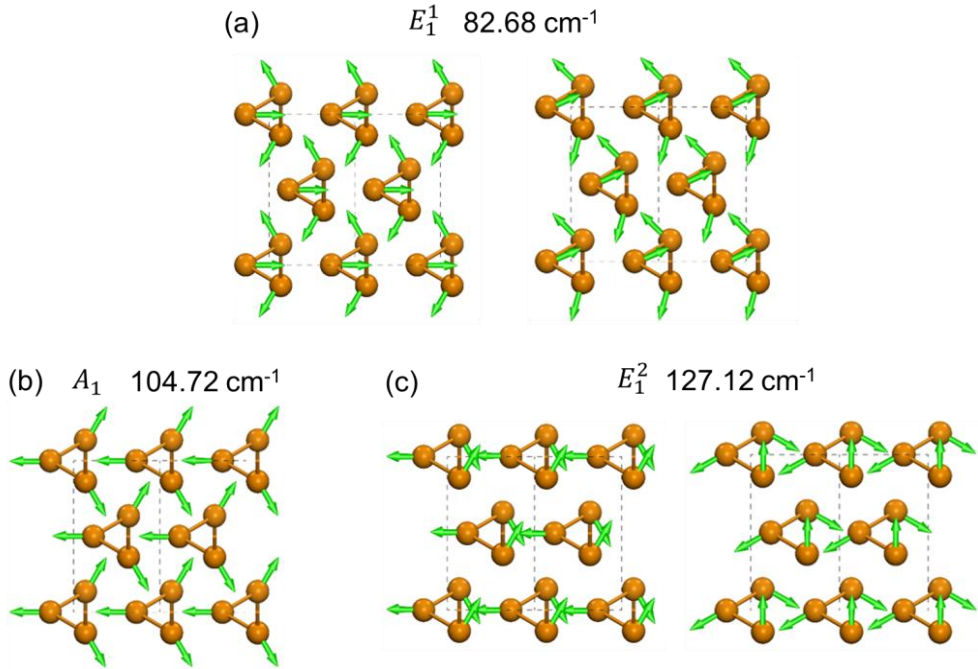

**Fig. S10.** Vibrational displacements for the three Raman activated optical modes of 3L-Te, i.e. $E_1^1$ (a), $A_1$ (b) and $E_1^2$ (c), of bulk-α-Te in a rectangular lattice.

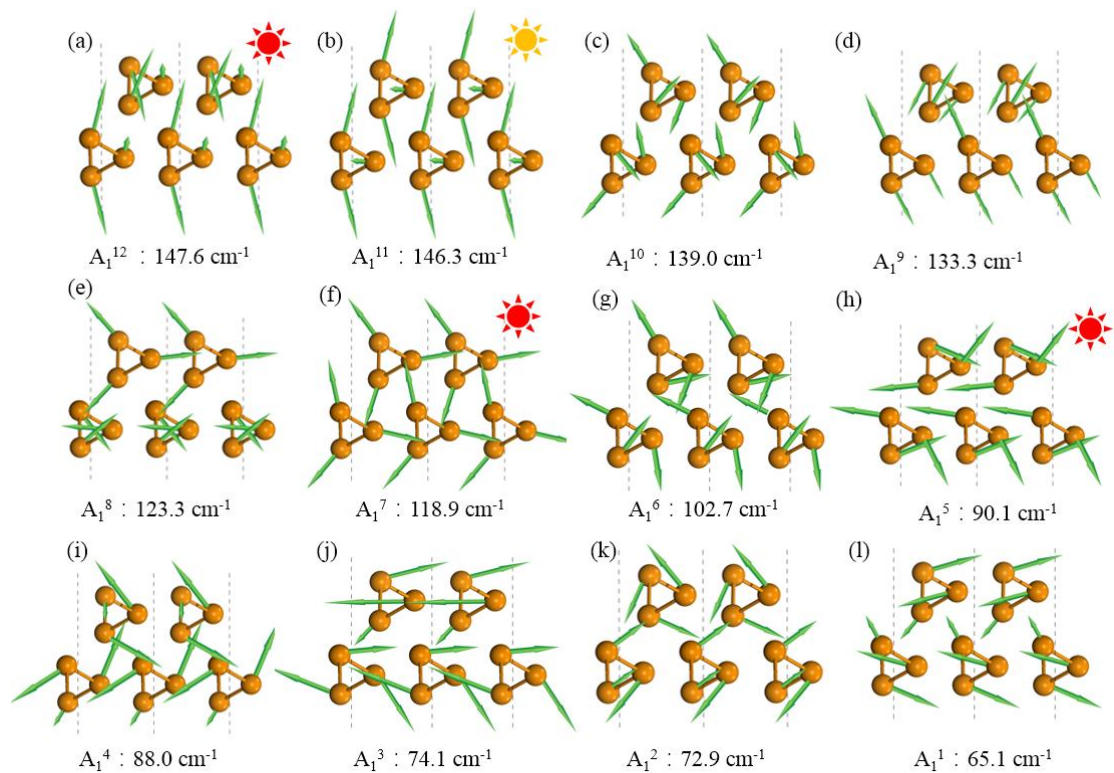

**Fig. S11.** Vibrational displacements for all DFPT-calculated Raman activated optical modes of 2L-α-Te. These three modes with the highest Raman intensities are marked by the red suns. Mode $A_1^{11}$ marked by the yellow sun is a pair mode of mode $A_1^{12}$.

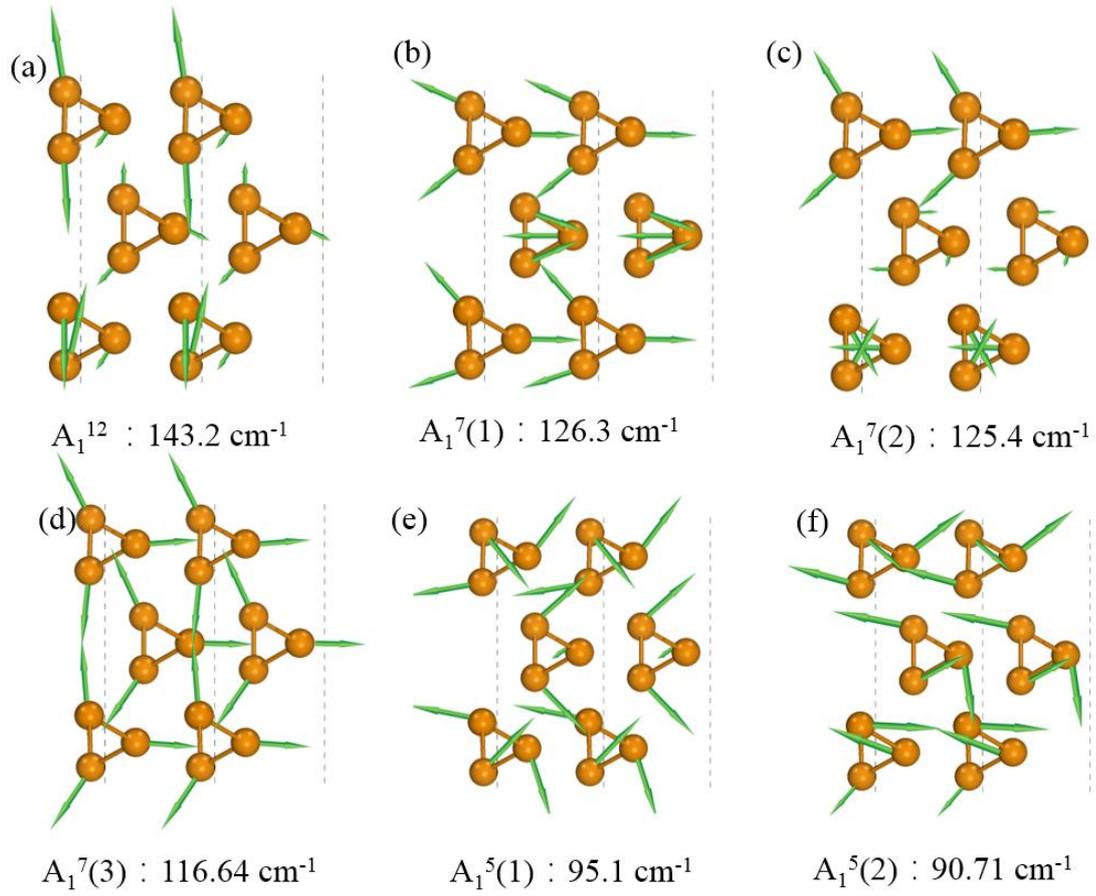

(a) $A_1^{12}$ : 143.2 cm$^{-1}$  (b) $A_1^7(1)$ : 126.3 cm$^{-1}$  (c) $A_1^7(2)$ : 125.4 cm$^{-1}$

(d) $A_1^7(3)$ : 116.64 cm$^{-1}$  (e) $A_1^5(1)$ : 95.1 cm$^{-1}$  (f) $A_1^5(2)$ : 90.71 cm$^{-1}$

**Fig. S12.** Vibrational displacements of six examples of the DFPT-calculated Raman activated modes of 3L-Te, $A_1^{12}$ (a), $A_1^7$ (b-d) and $A_1^5$ (e-f), which are distorted to mix with other modes at inner layers. Therein, modes $A_1^{12}$, $A_1^7$, and $A_1^5$ are mixed with modes $A_1^7$, $A_1^9$ and $A_1^6$, respectively.

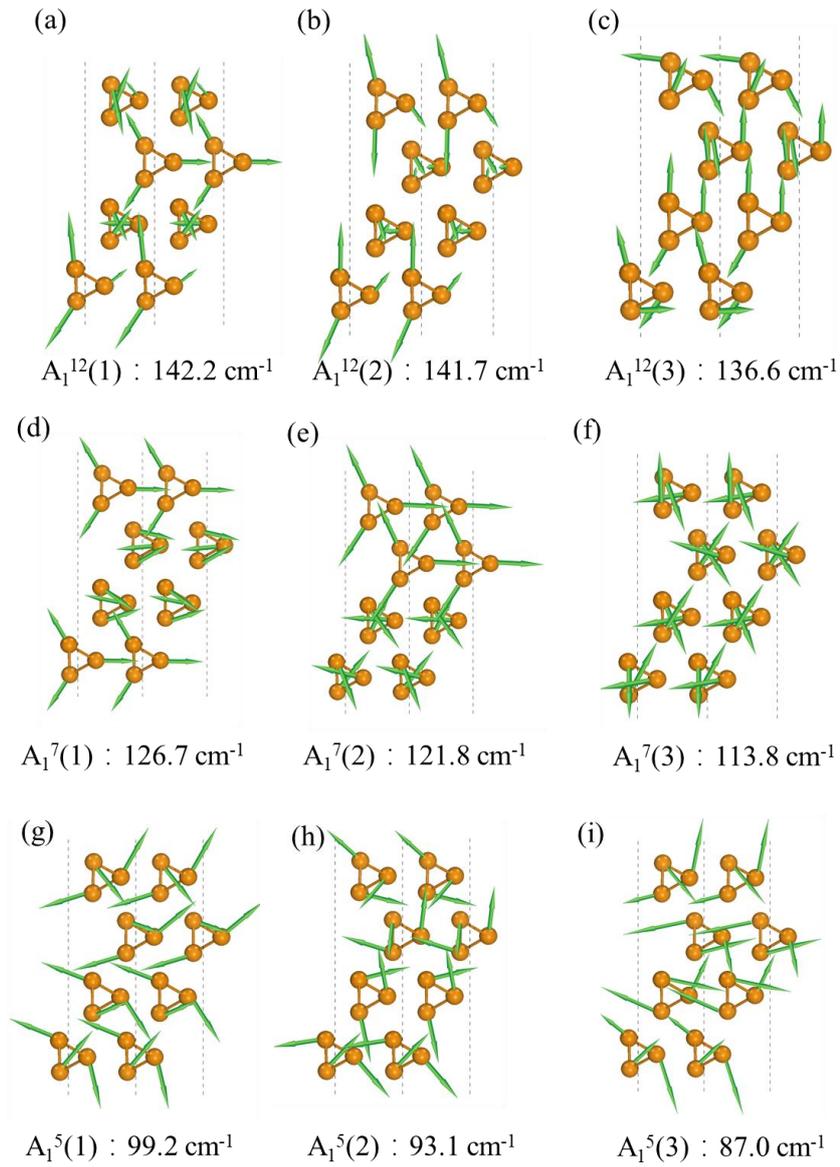

**Fig. S13.** Vibrational displacements of nine DFPT-calculated Raman activated modes of 4L-Te, $A_1^{12}$ (a-c), $A_1^{7}$ (d-f) and $A_1^{5}$ (g-i), which are distorted to mix with other modes at inner layers. Therein, mode $A_1^{12}$ is mixed with modes $A_1^{9}$ and $A_1^{7}$. Modes $A_1^{7}$ and $A_1^{5}$ are mixed with modes $A_1^{9}$ and $A_1^{6}$, respectively.

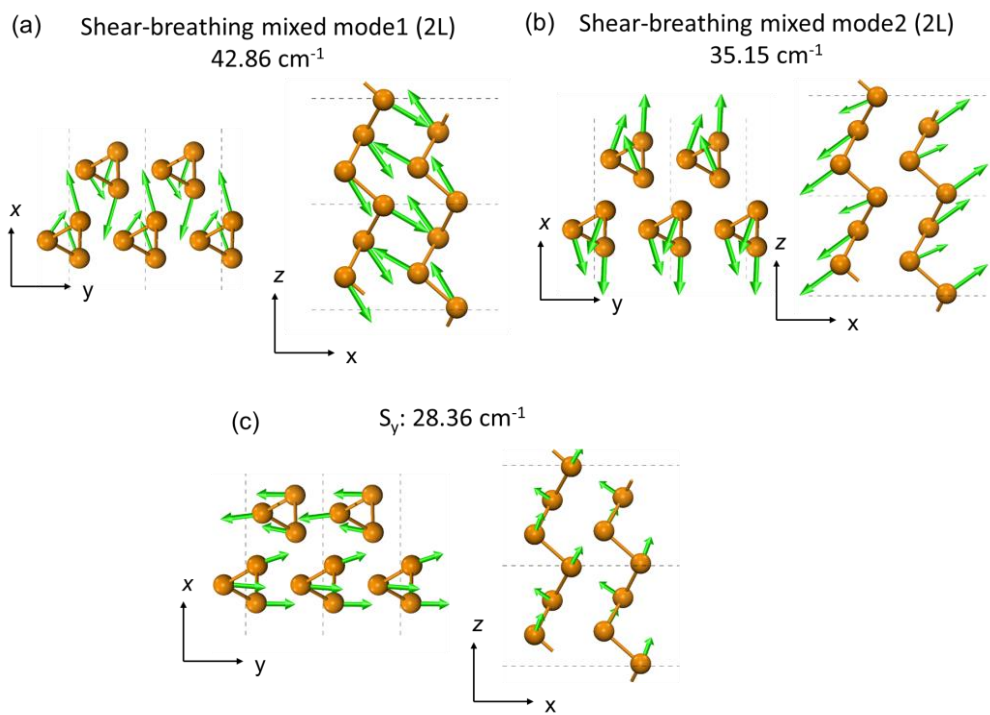

**Fig. S14.** Vibrational displacements of three DFPT-calculated low-frequency Raman activated interlayer acoustic modes of 2L-Te, namely two shear-breathing mixed modes (a-b) and shear mode along $y$ ($S_y$) (c).

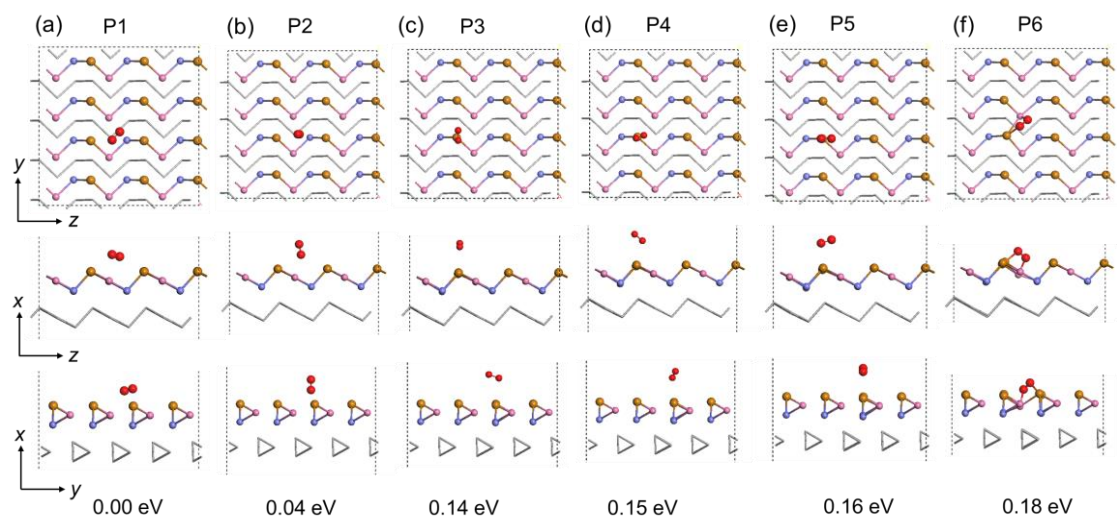

**Fig. S15.** Six representative stable atomic structures of physisorbed $O_2$ on 2L-α-Te among 27 $O_2$ physisorbed configurations. The P1 (a) configuration is the most stable adsorption site with -0.22 eV adsorption energy. The adsorption energies of configurations P2 (b), P3 (c), P4 (d), P5 (e) and P6 (f) are 0.04, 0.14, 0.15, 0.16 and 0.18 eV higher than that of the P1 configuration.

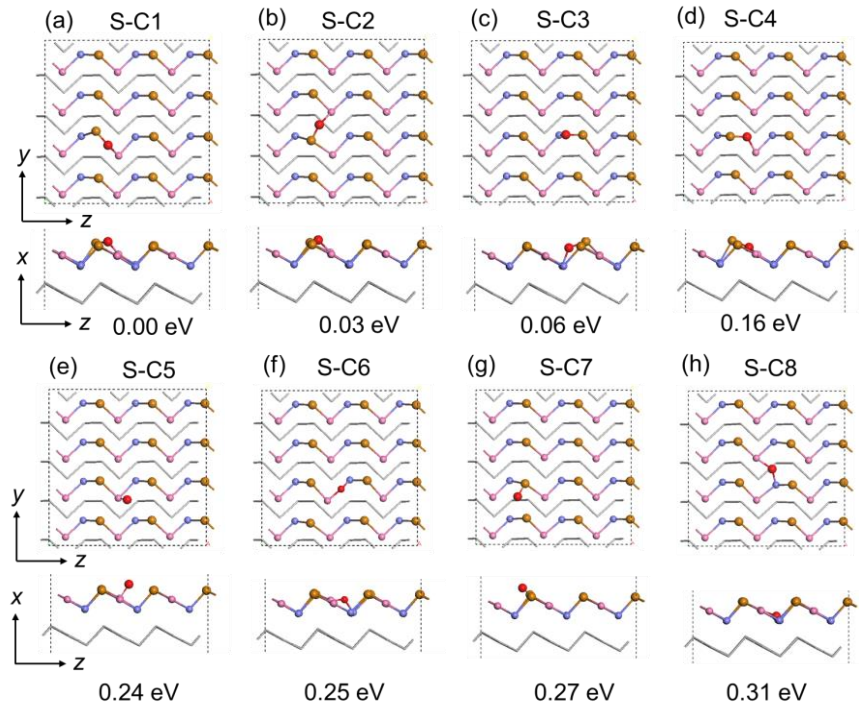

**Fig. S16.** All possible configurations of chemisorbed single O on 2L-α-Te. The S-C1 (a) configuration is the most stable adsorption site with an adsorption energy of -0.94 eV. The adsorption energies of configurations S-C2 (b), S-C3 (c), S-C4 (d), S-C5 (e), S-C6 (f), S-C7 (g) and S-C8 (h) are 0.03, 0.06, 0.16, 0.24, 0.25, 0.27 and 0.31 eV higher than that of the S-C1 configuration, respectively.

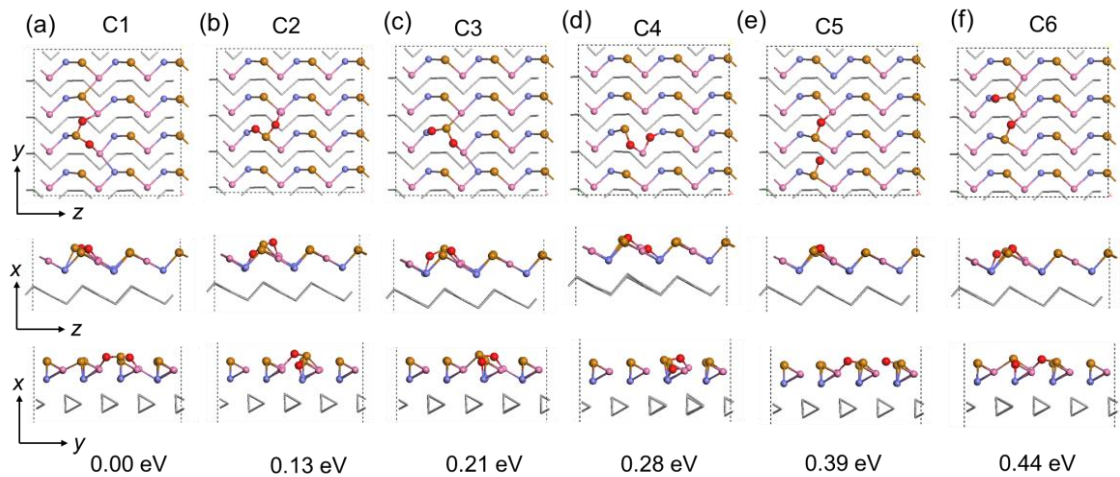

**Fig. S17.** Six representative stable atomic structures of chemisorbed $O_2$ on 2L-α-Te among 23 $O_2$ chemisorbed configurations. Configuration C1 (a) is the most stable adsorption site with a -2.31 eV adsorption energy. The adsorption energies of configurations C2 (b), C3 (c), C4 (d), C5 (e) and C6 (f) are 0.13, 0.21, 0.28, 0.39 and 0.44 eV higher than that of the C1 configuration, respectively.

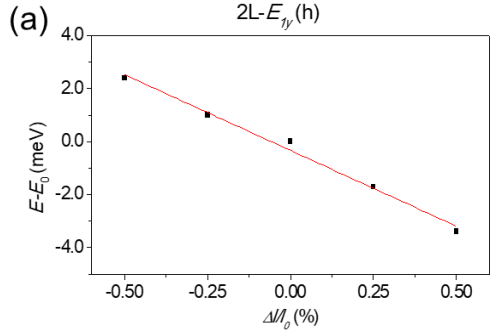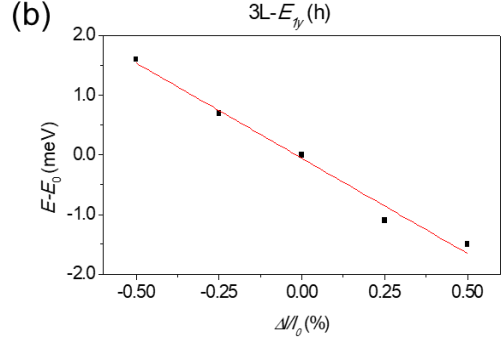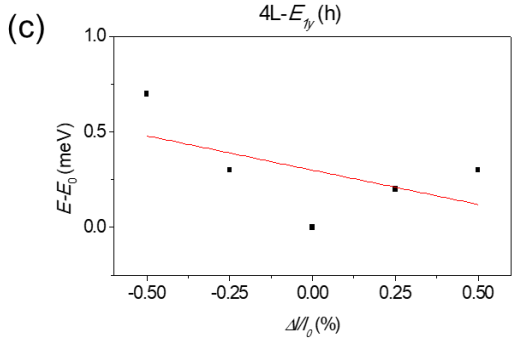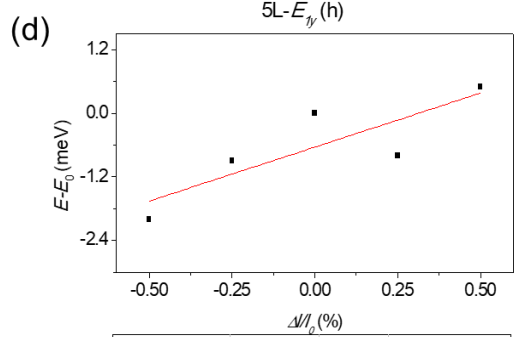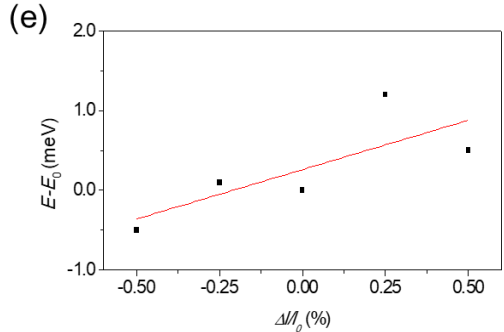

**Fig. S18.** Relative errors of the hole deformation potential along the *y* direction. Here, $E_{1y}$ (h) is derived from the changes of energy of the $i^{th}$ band under uniaxial along *z* from -0.5% ~ 0.5% with a step of 0.25%, which is the linear region of $E_{1y}$ (h).

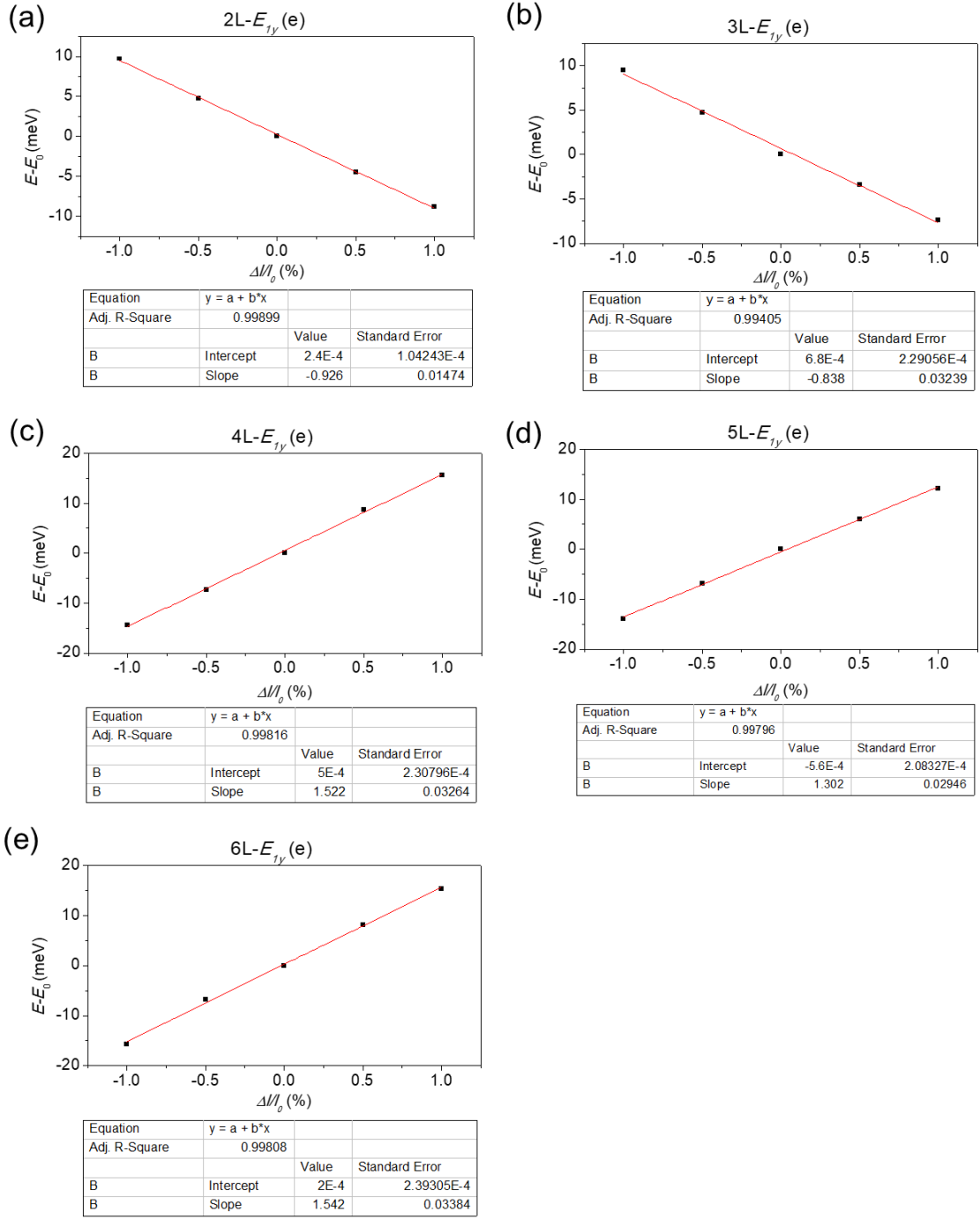

**Fig. S19.** Relative errors of the electron deformation potential along the *y* direction. Here, $E_{1y}$ (e) is derived from the changes of energy of the $i^{th}$ band under uniaxial along *y* from -1.0% ~ 1.0% with a step of 0.5%.

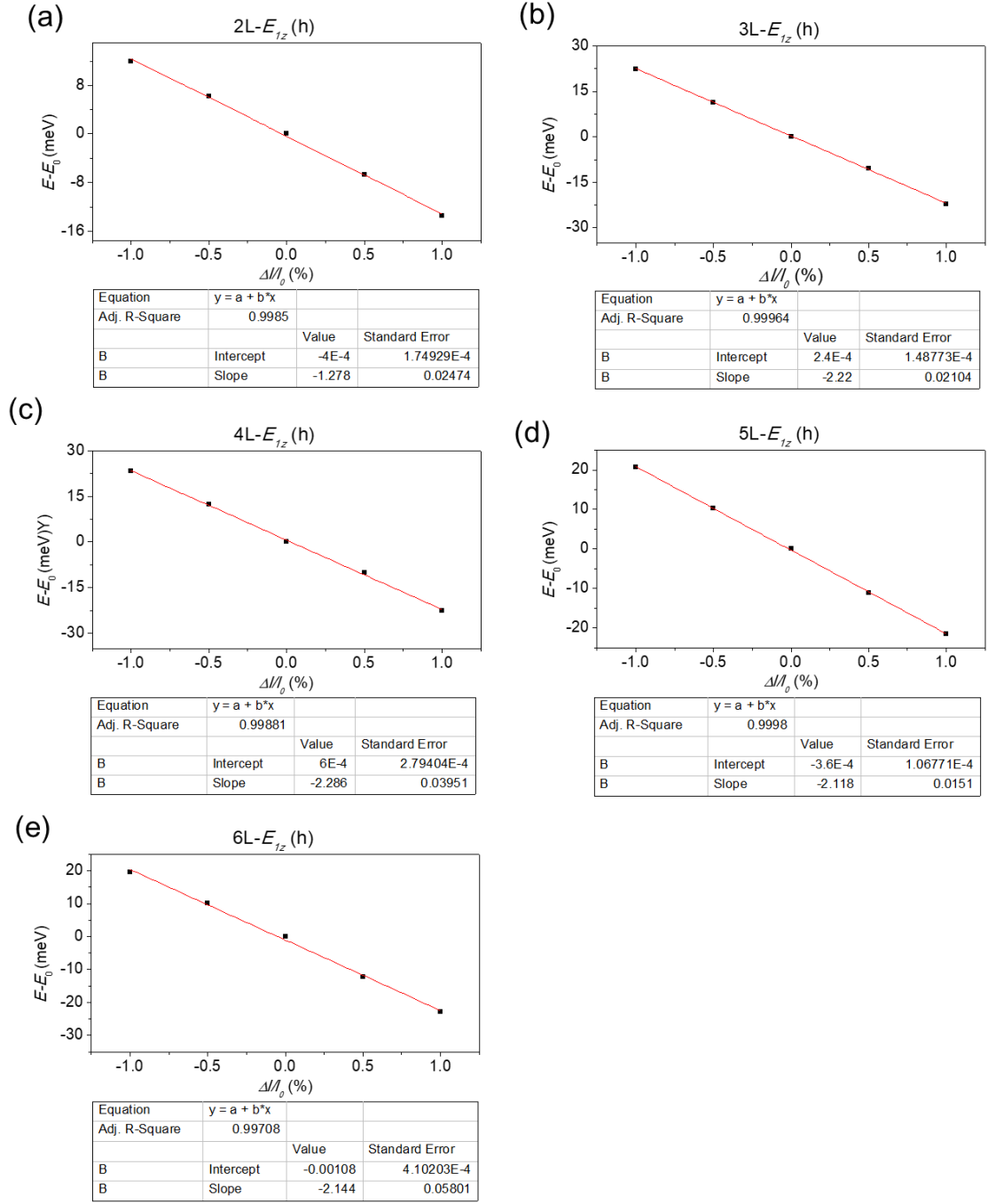

**Fig. S20.** Relative errors of the hole deformation potential along the $z$ direction. Here, $E_{1z}$ (h) is derived from the changes of energy of the $i^{th}$ band under uniaxial along $z$ from -1.0% ~ 1.0% with a step of 0.5%.

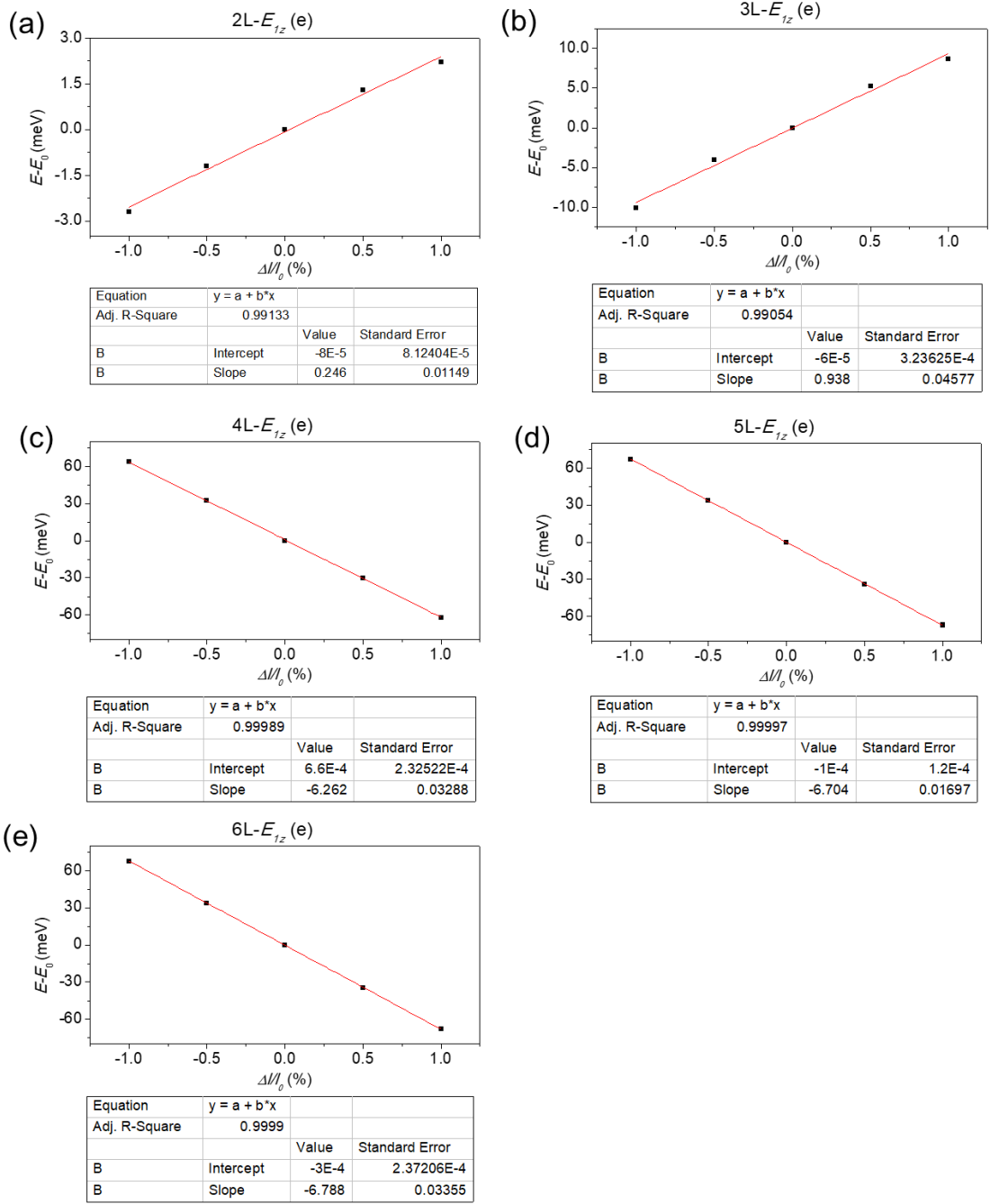

**Fig. S21.** Relative errors of the electron deformation potential along the $z$ direction. Here, $E_{1z}$ (e) is derived from the changes of energy of the $i^{th}$ band under uniaxial along $z$ from -1.0% ~ 1.0% with a step of 0.5%.

49